\documentclass[a4paper,usenatbib]{mnras}
\usepackage{graphicx}
\usepackage[fleqn]{amsmath}
\usepackage[T1]{fontenc}
\usepackage[varg]{txfonts}
\usepackage{color}
\usepackage{aecompl}
\usepackage{times}
\usepackage{ae,aecompl}
\usepackage{amssymb,amsfonts,hyperref,color}

\usepackage{hyperref}
\usepackage{bm}

\usepackage[encapsulated]{CJK}
\usepackage{ucs}
\usepackage[utf8x]{inputenc}
\newcommand{\cntext}[1]{\begin{CJK}{UTF8}{gbsn}#1\ignorespacesafterend\end{CJK}}  

\newcommand\be{\begin{equation}}
\newcommand\en{\end{equation}}

\title[Proto-Jupiter's Envelope]{Global 3D Radiation Hydrodynamic Simulations of Proto-Jupiter's Convective Envelope}

\author[Z.~Zhu et al.]{%
Zhaohuan Zhu \cntext{(朱照寰)}$^{1}$\thanks{E-mail: zhaohuan.zhu@unlv.edu}\,, Yan-Fei Jiang \cntext{(姜燕飞)}$^{2}$, Hans Baehr$^{1}$, Andrew N.  \newauthor Youdin$^3$, Philip J. Armitage$^{2,4}$, and Rebecca G. Martin$^{1}$ \\
$^{1}$Department of Physics and Astronomy, University of Nevada, Las Vegas, 4505 S.~Maryland Parkway, Las Vegas, NV~89154-4002, USA\\
$^{2}$Center for Computational Astrophysics, Flatiron Institute, 162 Fifth Avenue, New York, NY 10010, USA\\
$^{3}$Steward Observatory \& Department of Astronomy, University of Arizona, 933 N Cherry Ave, Tucson, AZ 85721, USA\\
$^{4}$Department of Physics and Astronomy, Stony Brook
  University, Stony Brook, NY 11794-3800, USA\\
\\
}

\date{In original form \today}

\begin{document}
\label{firstpage}
\pagerange{\pageref{firstpage}--\pageref{lastpage}} \pubyear{2021}
\maketitle

\begin{abstract}
The core accretion model of giant planet formation has been challenged by the discovery of recycling flows between the planetary envelope and the disc that can slow or stall envelope accretion. We carry out 3D radiation hydrodynamic simulations with an updated opacity compilation to model the proto-Jupiter's envelope. To isolate the 3D effects of convection and recycling, we simulate both isolated spherical envelopes and envelopes embedded in discs. The envelopes are heated at given rates to achieve steady states, enabling comparisons with 1D models. We vary envelope properties to obtain both radiative and convective solutions. Using a passive scalar, we observe significant mass recycling on the orbital timescale. For a radiative envelope, recycling can only penetrate from the disc surface until $\sim$0.1-0.2 planetary Hill radii, while for a convective envelope, the convective motion can ``dredge up'' the deeper part of the envelope so that the entire convective envelope is recycled efficiently. This recycling, however, has only limited effects on the envelopes' thermal structure. The radiative envelope embedded in the disc has identical structure as the isolated envelope. The convective envelope has a slightly higher density when it is embedded in the disc. We introduce a modified 1D approach which can fully reproduce our 3D simulations. With our updated opacity and 1D model, we recompute Jupiter's envelope accretion with a 10 $M_{\oplus}$ core, and the timescale to runaway accretion is shorter than the disc lifetime as in prior studies.  {Finally, we discuss the implications of the efficient recycling on the observed chemical abundances of the planetary atmosphere (especially for super-Earths and mini-Neptunes).}
\end{abstract}


\section{Introduction}
The core accretion mechanism is one of the leading giant planet formation mechanisms \citep{Perri1974,Mizuno1978}. 
It was developed using a combination of static 1D models and quasi-static evolutionary models. In the static models, the planet's envelope
structure is calculated for a given core mass given some assumptions on the envelope's thermal structure \citep{Perri1974, Mizuno1978, Mizuno1980}.  Assuming a luminosity that is released by a constant rate of planetesimal accretion \citep{Mizuno1980}, such static solutions suggest a maximum core mass for any given opacity. This maximum core mass
is the ``critical core mass'', beyond which the envelope is subject to collapse. Although the critical core mass can range widely
depending on the given luminosity and disc parameters \citep{Rafikov2006}, it is estimated to be around 10~$M_{\oplus}$ for Jupiter formation 
in the Solar Nebula \citep{Mizuno1980,Stevenson1982}. In assuming that all the luminosity comes from planetesimal
accretion, static models ignore energy release from the envelope's Kelvin-Helmholtz (KH) contraction.  Thus, these models implicitly assume that the envelope 
settles on a timescale much shorter than the core building timescale or the disc lifetime. 
We can think  of the static models as taking snapshots during the envelope accretion. To properly take into account
the envelope's KH contraction, quasi-static evolutionary models have been developed.  These models connect different
snapshots using energy conservation \citep{Bodenheimer1986, Pollack1996, Movshovitz10}. These quasi-static models suggest that Jupiter's formation has three stages: the core building stage, the atmosphere accretion stage, and the run-away stage when the core mass is comparable to the envelope mass. Run-away accretion ends when the planets manage to induce gaseous gaps in protoplanetary discs \citep{Bryden99,Alibert2005,Rosenthal2020}. The quasi-static approach has been extended to model the formation of giant planets in exoplanetary systems (e.g. \citealt{Piso2014,Piso2015, Lee2014, Lee2015, AliDib2020, Chen2020}). 

The adequacy of any 1D description of giant planet formation can reasonably be questioned. In addition to convection (as shown in 3-D simulations of \citealt{Ayliffe2012}) -- which can be modeled in 1D, but only approximately -- 3D numerical simulations show that the flow pattern around embedded planets is highly complex. Gas in the disc flows to the planet from the pole, and then leaves the planet from the midplane \citep{Machida2008, Tanigawa2012}. Isothermal \citep{Bate2003, Fung2015,Ormel2015,Bethune2019}, isentropic \citep{Fung2017}, adiabatic \citep{Fung2019} and radiation \citep{Paardekooper2008,Ayliffe2009,DAngelo2013,Szulagyi2016,Szulagyi2017,Cimerman2017,Lambrechts2017} hydrodynamic models all show complicated 3D flow patterns, but 
the details of the flow patterns are dramatically different among these simulations. In part, the differences arise due to different adopted equations of state (EOS) \citep{Fung2019}. Isothermal simulations show a rotationally-supported circumplanetary disc \citep{Tanigawa2012,Wang2014,Fung2019}, while adiabatic and radiative simulations show a pressure supported sphere. The effect of these 3D flow patterns on giant planet atmosphere accretion is still unclear. Some works suggest that significant ``atmosphere mass recycling'' \citep{Ormel2015} extends all the way to the planetary core or the simulated inner boundary \citep{Cimerman2017,Bethune2019}. This recycling could stall atmosphere accretion, preventing the planet's run-away accretion \citep{Moldenhauer2021}. Other works suggest that the bound inner envelope is not affected strongly by recycling \citep{Lambrechts2017,Fung2019}.

The 1D and 3D approaches have quite different assumptions and produce different results. Both approaches have shortcomings. Although 3D simulations are needed to capture recycling flows, they suffer from limited spatial resolution and a limited timespan. It is impractical to simulate Jupiter all the way from the core to the 5~au scale over the 1~Myr KH contraction timescale. Thus, 3D simulations normally place the inner boundary condition at radii larger than the physical radius of the core \citep{Bethune2019}, and often reduce the opacity by several orders of magnitude to shorten the KH timescale \citep{Cimerman2017, Moldenhauer2021}. A promising approach is to combine 3D simulations with 1D models. We can study detailed physical processes and measure key quantities from 3D simulations, and then use these 3D simulations to construct 1D models to study the long-term planet evolution.  

Here, we present the results of new 3D simulations of the forming Jupiter's convective envelope. The simulations use an updated opacity, appropriate for both disc and envelope physical conditions, and a radiative transfer scheme that directly solves the time-dependent transfer equation for the specific intensity \citep{Jiang2014}. This approach is more accurate than flux-limited diffusion at low and intermediate optical depths. In order to systematically isolate 3D effects due to convection and recycling, and to compare 1D and 3D results, our simulation setup is distinct from prior work in two ways. First, we inject  specified luminosities throughout the envelopes so that the envelopes can reach steady states. 
The steady-state outcome is similar to the traditional static 1D approach, making comparison between 1D models and 3D simulations easier. Second, we run 3D simulations of isolated spherical envelopes as well as envelopes that are embedded with the disc. The isolated 3D simulations have identical setups as 1D models, except that convection is simulated self-consistently. By comparing these 3D simulations with 1D radiation-only simulations, we are able to study the effects of convection on the envelope structure. At the same time, the 3D isolated envelope simulations are also similar to the 3D envelopes that are embedded in discs, except for the absence of disc recycling. A comparison between these two types of simulation can therefore quantify the effects of recycling.


The plan of the paper is as follows. We introduce our updated opacity in Section \ref{sec:opacity} before presenting the simulation methods in Section \ref{sec:method}. We present our results in Section \ref{sec:results}. After some discussion in Section \ref{sec:discussion}, we conclude in Section \ref{sec:conclusion}. We defer some numerical details to the Appendices.

\section{Opacity}
\label{sec:opacity}

Opacity determines the envelope's cooling ability and thus directly controls the envelope's KH contraction and later run-away accretion. Although we know little empirically about the opacity in the planetary envelope itself \citep[which will be affected by pebble accretion and coagulation processes:][]{Podolak03,Brouwers2020},
the outer envelope is directly connected with the protoplanetary disc, and we have constraints on the disc's dust opacity from ALMA observations.  For the deeper parts of the envelope where molecular and atomic opacities dominate, several widely used opacities (e.g. the opacities adopted in the MESA code \citealt{Paxton2011}) are not suitable for the conditions of the forming planet's envelope. Thus, we update and generate a new opacity table including dust, molecular, and atomic opacities, spanning a large density and temperature range for our envelope simulations. The Rosseland and Planck mean opacity table can be downloaded at the Github repository\footnote{\url{https://github.com/zhuzh1983/combined-opacity}}.

\subsection{Dust Opacity}
We calculate the dust opacity following \cite{Birnstiel2018}.  They compared several widely used dust opacities, and calculated a new set of opacities which have been successfully used in the ALMA large program DSHARP \citep{Andrews2018}. 
The dust is composed of four different materials: water ice \citep{Warren2008}, astronomical silicates \citep{Draine2003}, troilite, and refractory organics \citep{Henning1996}. The mass fraction of water ice, silicates, troilite, and refractory organics is 0.2, 0.3291, 0.0743, 0.3966 respectively \citep{Birnstiel2018}. These fractions are not derived from chemical equilibrium calculations.
Instead, they are determined based on protoplanetary disc SED constraints (e.g. the adoption of the low water abundance). 
Thus, the elemental abundance of the dust component is different from the solar abundance that is used for the molecular and atomic opacity calculations in the following sections. Although chemical equilibrium calculations for the dust components can remedy this inconsistency \citep{Ferguson2005}, we decide to use the dust components that are more consistent with protoplanetary disc observations.  
We adopt the nominal $q=3.5$ power law dust size distribution with a minimum particle size of 10$^{-5}$ cm. We have calculated the opacity for different maximum particle sizes: $a_{\rm max}$=10 $\mu$m, 1 mm, 1 cm, and 10 cm.   
The details of the opacity calculation can be found
in \cite{Birnstiel2018}\footnote{\url{https://github.com/birnstiel/dsharp_opac}}.

With increasing temperature, the four materials considered sublimate one after another. We adopt the dust sublimation temperatures in Table 3 of \cite{Pollack1994} to decide which material to remove at different temperatures. The sublimation temperatures of the water ice and refractory organics are from ``Water ice in disks'' and "Refractory organics in disks" in \cite{Pollack1994}. When the gas density is beyond the density range provided by  \cite{Pollack1994}, which is between 10$^{-18}$ to 10$^{-4}$ g cm$^{-3}$, we just use the limiting density (10$^{-18}$ or 10$^{-4}$ g cm$^{-3}$). 
Since dust gradually sublimates over a temperature range around the sublimation temperature, we smooth out the opacity transition around the dust sublimation temperature ($T_{\rm sub}$) using a sin function, 
\begin{eqnarray}
\kappa(T)=\kappa_{lT} + (\kappa_{hT}-\kappa_{lT})\left(1-\frac{1}{2}\left({\rm sin}\left(\frac{T-T_{\rm sub}}{\Delta T_{\rm sub}}\pi\right)+1\right)\right)\\\nonumber
{\rm if}\; T_{\rm sub}-\frac{\Delta T_{\rm sub}}{2}<T<T_{\rm sub}+\frac{\Delta T_{\rm sub}}{2}\,,
\end{eqnarray}
where $\kappa_{lT}$ and $\kappa_{hT}$ are opacities before and after one particular material is sublimated. To determine the width of the transition temperature ($\Delta T_{\rm sub}$), we use
the GRAINS code \citep{PETAEV2009, Li2020} and Figure 4 in \cite{Li2020} to estimate $\Delta T_{\rm sub}=T_{\rm sub}/10\times(\log_{10}(P/{\rm bar})/16+1)$ where $P$ is the gas pressure. The pressure dependent term accounts for the fact that the transition is smoother in a higher pressure environment.

\begin{figure}
\includegraphics[width=3.8in]{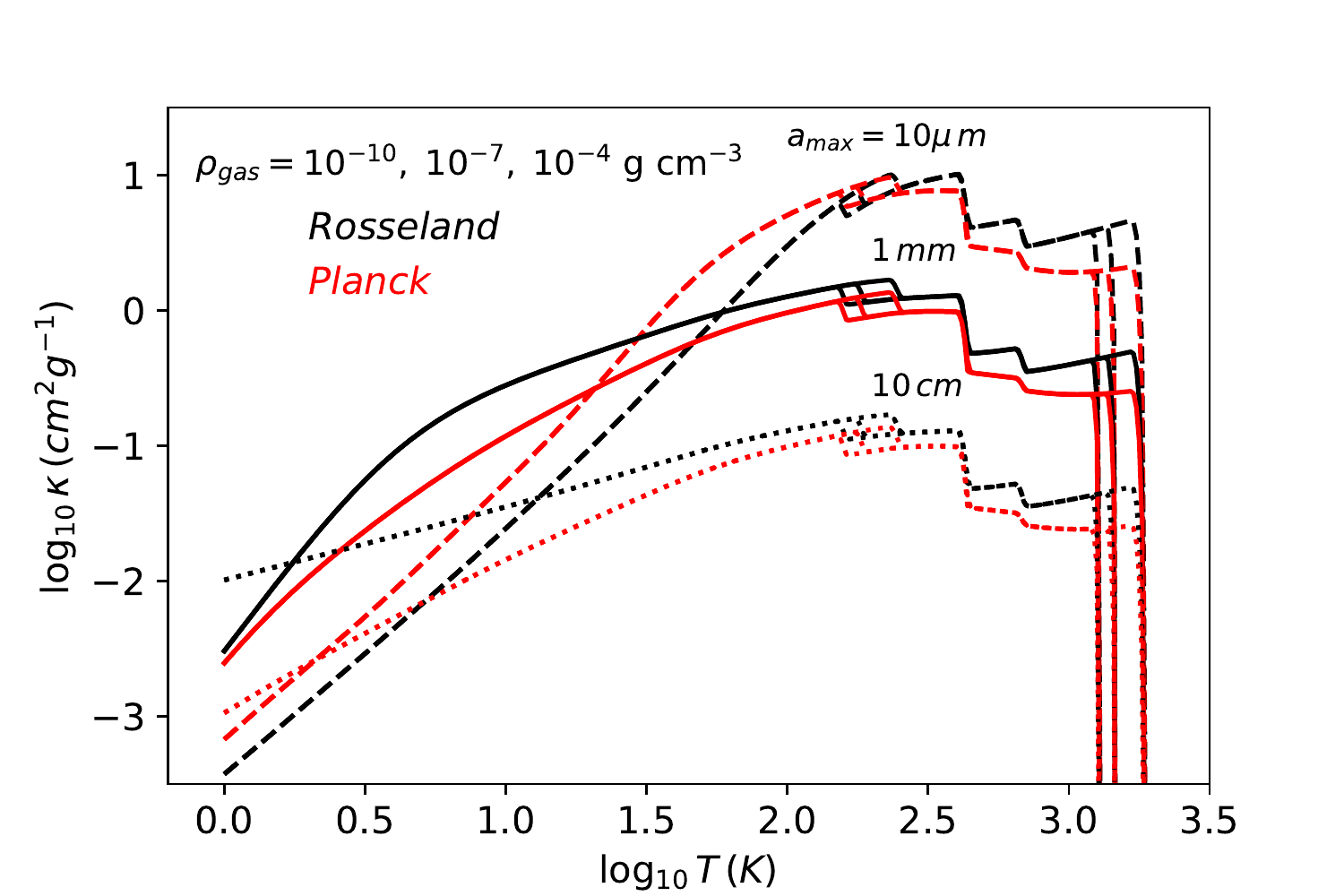}
\caption{The Rosseland mean (black curves) and Planck mean (red curves) opacities for dust with different maximum particle size, $a_{\rm max}$ (three different line types: dashed, solid, and doted curves). 
To show that the dust sublimation temperature depends on the gas density,
there are three different
curves (almost overlapping each other) for each type that are calculated under three different gas densities ($10^{-10}$, $10^{-7}$, and $10^{-4}$ g cm$^{-3}$).    }
\label{fig:rossplanck}
\end{figure}

The Rosseland mean and Planck mean opacities for dust with different $a_{\rm max}$ are shown in Figure \ref{fig:rossplanck}. When the temperature increases,
different dust components start to sublimate. The sublimation occurs at a higher temperature in a higher pressure environment, except for troilite and refractory organics whose sublimation temperature was given as a single value in \cite{Pollack1994}.

\begin{figure*}
\includegraphics[trim=25mm 8mm 6mm 6mm, clip, width=7.6in]{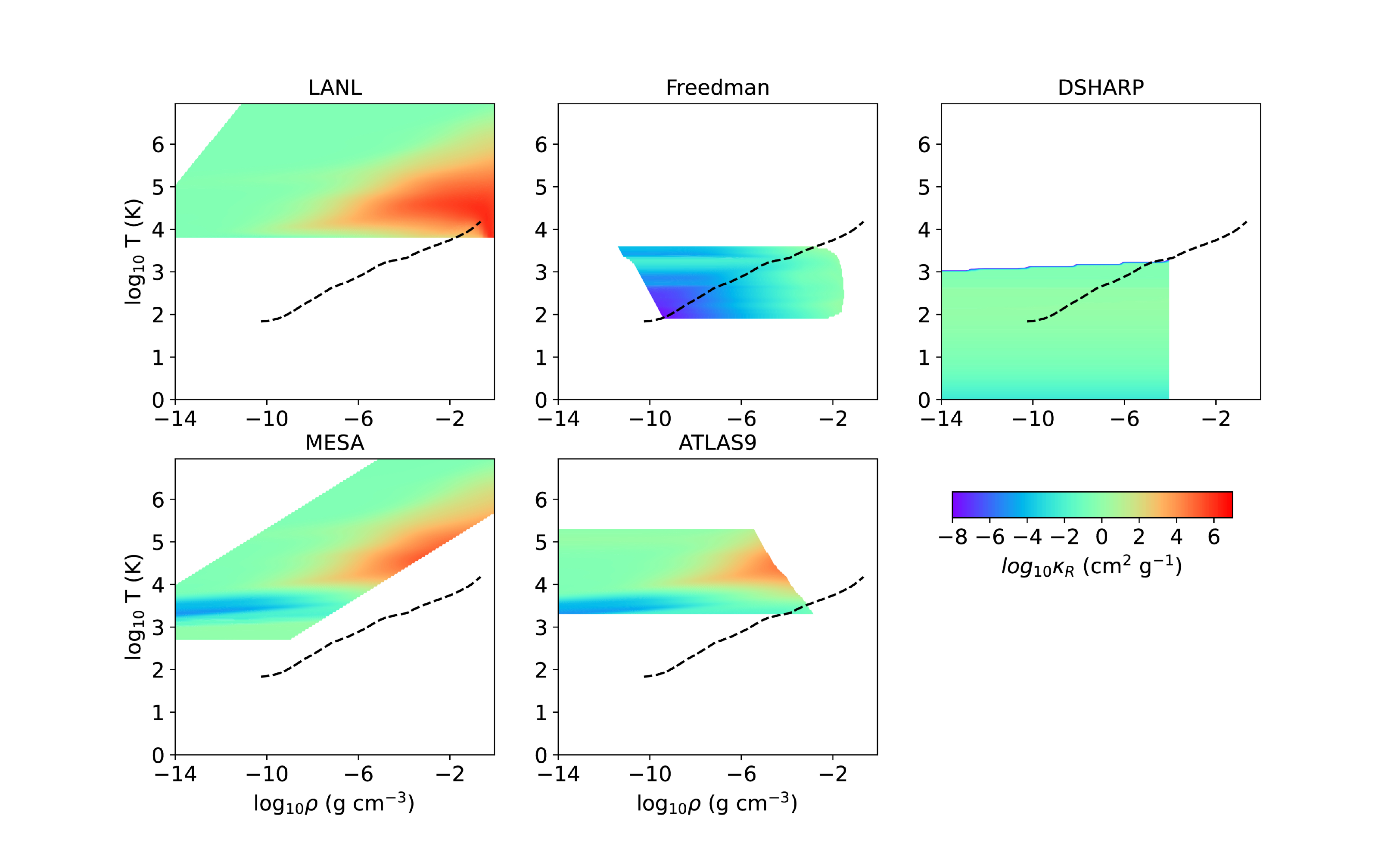}
\caption{The Rosseland mean opacities at different densities and temperatures from various databases. The colored region
is where the data are calculated and valid. The dashed curve labels the $\rho-T$ condition of a forming Jupiter's envelope at the cross-over mass, as calculated in Section \ref{sec:jupiterevo}. 
 \label{fig:opacitycompare}}
\end{figure*}

\subsection{Molecular Opacity}
\label{sec:moleopa}
For the gas opacity at low temperature (dominated by the molecular opacity), we adopt the opacity table provided by \cite{Freedman2014}\footnote{\url{https://www.ucolick.org/~jfortney/models.htm}}. This opacity table covers gaseous pressure
from $10^{-6}$ to 300 bar, and temperature from 75 to 4000 K. The corresponding gaseous density range is from $\sim 10^{-10}$ to $10^{-3}$ g cm$^{-3}$ (Figure \ref{fig:opacitycompare}), which is significantly
broader than the density/pressure range in previous gaseous opacity tables (e.g. \citealt{Freedman2008}, \citealt{Ferguson2005} ). On the other hand, deep within the forming planet's envelope, the gaseous density
can be much higher, reaching 1 g cm$^{-3}$. Thus, we use the analytical fit provided in \cite{Freedman2014} (Equations 3-5) to derive the Rosseland mean opacity for densities higher than
$\sim 10^{-3}$ g cm$^{-3}$ and lower than 1 g cm$^{-3}$.  Applying the analytical fit to high densities leads to a smooth transition to atomic opacities in the next section. However, we caution that there is no detailed opacity calculation to show that such an extension of the analytical fit to  high density is valid. For Planck mean opacities, the analytical fit formula has not been provided. Thus, we simply assume that the Planck mean opacity at densities beyond the table range in \cite{Freedman2014} is the same as the Planck mean opacity at the highest density in the table at that temperature. For opacity at densities lower than the provided table range, we also assume that the opacity is the same as the opacity at the lowest density in the table.

\subsection{Atomic Opacity}
At higher temperatures where molecules have been dissociated and the atomic opacity dominates, we adopt the 
LANL new generation opacity table \citep{Colgan2016}\footnote{\url{https://aphysics2.lanl.gov/apps/}}. This new LANL opacity table is LANL's recent opacity effort 
using their ATOMIC opacity and plasma modeling code \citep{Magee2004,Hakel2006} with their atomic data \citep{Fontes2015}.  It provides opacities for elements from hydrogen through zinc
for a wide range of temperatures (0.5 eV up to 100 kev) and densities (spanning at least 12 orders of magnitude). To derive the solar opacity, 
we use the solar elemental abundances provided by \cite{Asplund2009} (photosphere abundance in their Table 1). To calculate the opacities at the metal rich/poor environment, we vary the metal (beyond H and He) abundance accordingly. 
Both Planck mean and Rosseland mean opacities have been generated
using the LANL opacity website. Since protoplanetary discs and planet envelopes are cooler and less dense than stars, we only calculate the opacity with temperatures from 0.5 eV  to 1 keV 
($\sim$5800 K to 10$^7$ K) and densities from 10$^{-14}$ to 1 g cm$^{-3}$. The density range where the LANL opacity is valid depends on the temperature and element considered. 

One advantage of the LANL opacity table over the widely used opacities (mainly from \citealt{Ferguson2005} and \citealt{Iglesias1996}) in MESA \citep{Paxton2011} is that it includes opacities at high densities and low temperatures, the conditions found in the planet envelope/atmosphere. 
For example, the MESA opacity table has a maximum opacity at $\rho\sim0.01$ g cm$^{-3}$ and $T\sim3\times10^4$ K (Figure 3 in \citealt{Paxton2011}). 
This maximum opacity is artificial since the adopted radiative opacity in MESA is only valid
for $\rho\lesssim0.01$ g cm$^{-3}$ at $T\sim3\times10^4$ K, as shown in Figure \ref{fig:opacitycompare}. For $\rho\gtrsim$0.01 g cm$^{-3}$, it is interpolated to connect to the electron conduction opacity \citep{Cassisi2007} at much higher densities. 
Considering plasma screening and electron degeneracy effects (e.g. Pauli blocking, \citealt{ARMSTRONG201461}), the LANL radiative opacity is valid at much higher densities, covering the high pressure range that we are interested in (Figure \ref{fig:opacitycompare}). 
By comparison, we can see that, for a given temperature, the LANL radiative opacity continues to rise with density even to our upper density limit of $\rho=1 $g cm$^{-3}$, which is dramatically different from the MESA opacity.

\begin{figure*}
\includegraphics[trim=20mm 90mm 6mm 6mm, clip, width=7.in]{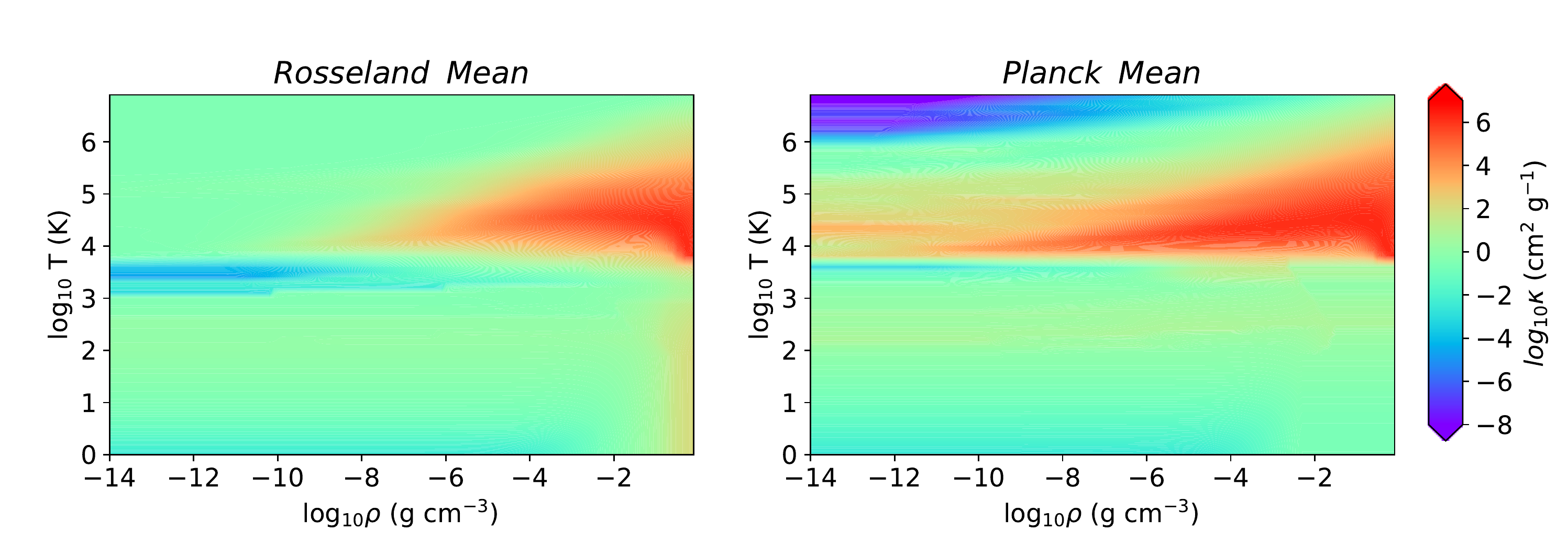}
\caption{ The final combined Rosseland mean and planck mean opacities. 
 \label{fig:combine}}
\end{figure*}

\begin{figure}
\includegraphics[width=3.6in]{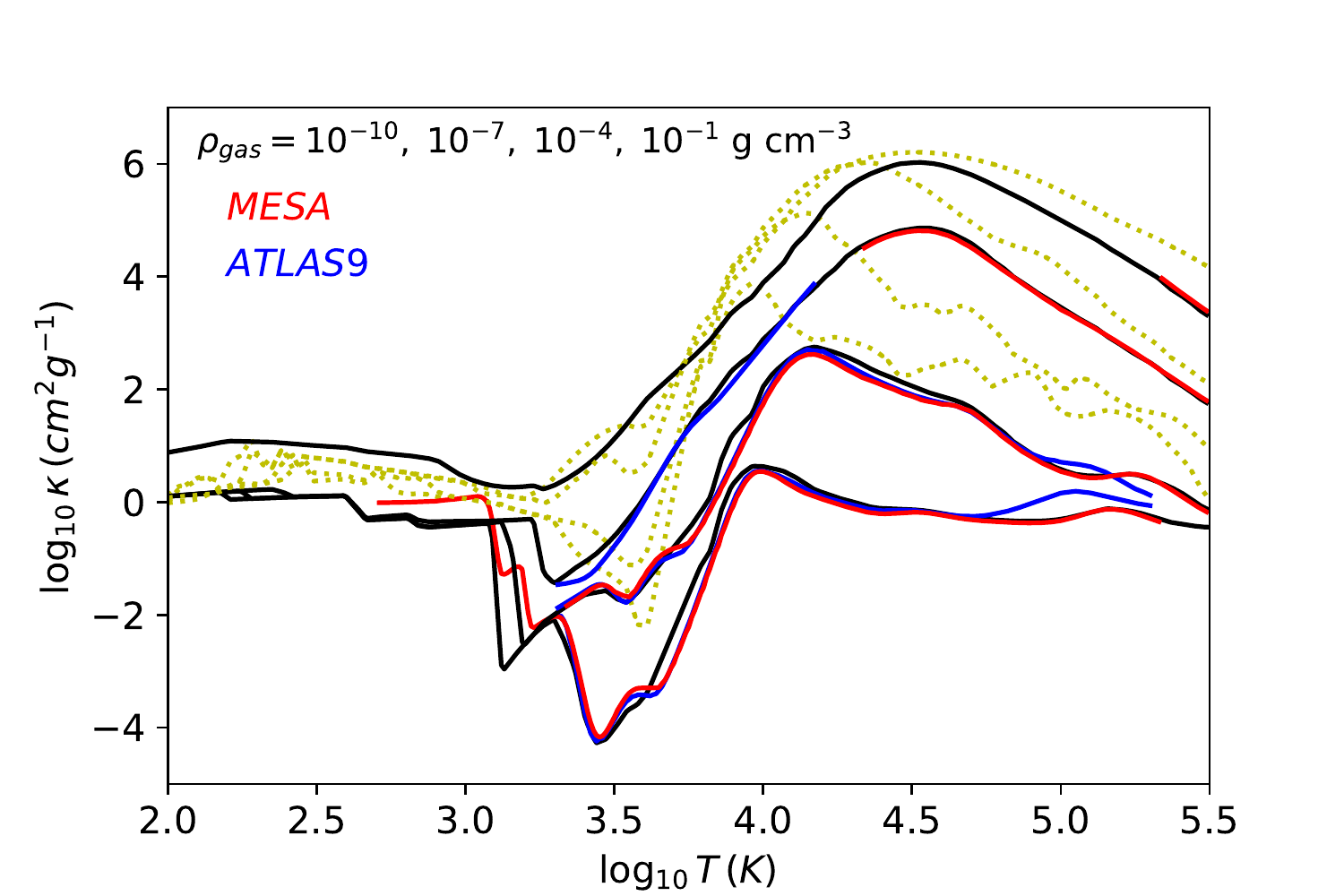}
\caption{Our combined Rosseland mean (black curves) and Planck mean (yellow dotted curves) opacities compared with the MESA and ATLAS9 Rosseland mean opacities
at different gas densities and temperatures. Only the opacities at their valid $\rho-T$ conditions are plotted.
In the common $\rho-T$ parameter space where different opacity databases are all valid, our opacities agree with the other opacities quite well.   }
\label{fig:rossplanckcombine}
\end{figure}

\subsection{Combining Opacities}

Various opacities from different opacity databases, together with their valid range, are shown in Figure \ref{fig:opacitycompare}.  The top three panels
show the atomic, molecular, and dust opacities we adopt. The bottom two panels show the solar abundance opacities used in MESA \citep{Paxton2011} and ATLAS9 \citep{Castelli2003}. Since MESA and ATLAS9 are stellar structure codes, their adopted opacities miss the $\rho-T$ condition of the planet interior. 

Our combined opacities (LANL+Freedman+DSHARP) are shown in Figure \ref{fig:combine}. At 100 K $<$ T $<$1000 K, molecules and dust coexist and they all contribute to the total opacity. Since \cite{Freedman2008} didn't provide the monochromatic opacity, we cannot derive the total monochromatic opacity to calculate the mean opacities. Thus, we simply pick the maximum between the molecular and dust mean opacities as the total opacity. Considering that the molecular opacity is normally significantly smaller than the dust opacity, the derived opacity should approximate the real mean opacity reasonably well. 
For the $\rho-T$ region that does not have valid opacity data,
we  use the opacity at the smallest or largest available density at that specific temperature, except for the molecular opacity (the detailed interpolation scheme for the molecular opacity is discussed in Section \ref{sec:moleopa}). 

To verify our combined opacities, we compare our mean opacities with the MESA and ATLAS9 opacities in Figure \ref{fig:rossplanckcombine}. Our opacities are
valid over a much larger $\rho-T$ parameter space. In the common $\rho-T$ parameter space where the different opacity databases are all valid, our opacities agree with the other opacities quite well. These combined opacities are significant updates to our previously compiled opacities for protoplanetary disc numerical simulations \citep{Zhu2009b,Zhu2012a}.  After deriving this new opacity table, we are ready to use it in our numerical simulations.

\section{Numerical Method}
\label{sec:method}
We solve the hydrodynamical equations using \texttt{Athena++} \citep{Stone2020}.
\texttt{Athena++} is the successor of \texttt{Athena} that uses a higher-order Godunov scheme for MHD and 
the constrained transport (CT) to conserve the divergence-free property for magnetic fields \citep{GardinerStone2005,GardinerStone2008,Stone2008}.  
The geometric source terms in curvilinear coordinates 
(e.g. in cylindrical and spherical-polar coordinates) are specifically implemented to conserve angular momentum to  machine precision. 
This property is crucial for disc simulations. In this work, we further extend the angular momentum conservation property to the rotating frame, as detailed in Appendix \ref{sec:source}.

We have used two different radiation modules to solve the radiative transfer equation and to couple radiation with the fluid equations. Both modules directly solve the
time-dependent radiative transfer equation for the specific intensity \citep{Jiang2014}. Unlike the commonly used flux-limited diffusion approximation \citep{Levermore1981} and 
the M1 closure method \citep{Gonzalez2007,SkinnerOstriker2013}, solving the specific intensity equation does not suffer the shadowing problem or the ray crossing 
problem \citep{Davis2012}. 
We first use the method of \cite{Jiang2014} to explicitly solve the radiative transfer equation. Since the characteristic speed in the transport step is the speed of light, 
solving this equation explicitly requires very small numerical timesteps. Thus, we adopt the reduced speed of light approach
as in \cite{Zhang2018}. Although this explicit method is numerically stable and robust, the reduced speed of light could lead to inaccurate light crossing and diffusion timescales (e.g. \citealt{Zhang2018, Zhu2020}). Thus, after the simulation settles to a steady state (5 planetary orbits for 3-D simulations), we restart the simulations using the implicit method of \cite{Jiang2021}.  
Although the implicit method is slow due to the iteration steps and the speed of light is still reduced in some cases to speed up the convergence, we can achieve a much better timescale separation between the radiation transport and fluid dynamics. 
In most of our cases, the speed of light only needs to be reduced at most by a factor of 10 for the implicit method, compared with a factor of $2\times 10^5$ for the explicit method. 
On the other hand, using the explicit and the implicit methods  give almost identical results in our simulations, suggesting a good timescale separation even for the explicit method. Test problems using either the implicit and explicit methods are provided in Section \ref{sec:test}. 

We adopt the third-order piecewise parabolic method (PPM) for spatial reconstruction for hydrodynamic quantities, which maintains a better hydrostatic equilibrium compared with the second-order reconstruction method. For the explicit radiative transfer, we also adopt the third-order spatial reconstruction for the intensity in the radiation transport step, which is crucial for  accurately simulating regions with high optical depths (e.g. the optical depth of $10^4$, \citealt{Zhu2020}). For the implicit radiative transfer scheme, the second-order intensity reconstruction is sufficient for capturing the thermal structure of the optically thick region, and is thus adopted. We use the second-order Van-Leer method for the time integration, and the HLLC Riemann solver to calculate the flux for hydrodynamic quantities. 

\subsection{Protoplanetary Disc Structure and Planetary Core Properties}
\label{sec:diskstructure}
The protoplanetary disc structure determines the planet's atmosphere structure.  For example, the disc temperature directly affects the envelope crossover time \citep{AliDib2020}. Unlike most previous works \citep{Piso2014, Lee2014} which use the minimum-mass solar nebula (MMSN) or the minimum-mass extrasolar nebula model  (MMEN, \citealt{Chiang2013}), we adopt the disc structure from \cite{Alessio1998} which successfully fits protoplanetary disc observations. Its thermal  structure is also self-consistently derived with disc radiative transfer calculations.  We fit Figure 3 of \cite{Alessio1998} to derive the disc midplane temperature as
\begin{equation}
T_c=\begin{cases}
39.4\times (R/10\, au)^{-4/5}\; {\rm K} \;\;\;\; if\;\;\;R<10\;{\rm au}\\
39.4  \times (R/10\, au)^{-1/2}\; {\rm K} \;\;\;\; if\;\;\;R>10\;{\rm au}\,.\label{eq:Tc}\\
\end{cases}
\end{equation}
In the disc region beyond 10 au, stellar irradiation dominates. 
As in \cite{Alessio1998}, we derive
the disc surface density using the $\alpha$ disc model \citep{Shakura1973}. We also require
the disc to be gravitationally stable.
Thus, we have surface density $\Sigma=min\{\dot{M}/(3\pi\nu),\Sigma_{Q=1}\}$, where $\nu=\alpha c_{s}^2/\Omega$
and $\Sigma_{Q=1}=c_{s}\Omega/(\pi G)$. We choose a constant accretion rate through the disc $\dot{M}=10^{-8}M_{\odot}/yr$ and $\alpha=10^{-3}$ for our disc model. 
Thus, at the disc radius $R$=5 AU, the disc's surface density is 488 g/cm$^2$, the midplane temperature is 66 K, the midplane density is 7$\times 10^{-11}$ g/cm$^3$, and the
disc's aspect ratio (H/R) is 0.037. With similar assumptions, our adopted disc model is similar to the \cite{Bitsch2015} model.  

We choose 10 $M_{\earth}$ as the planet's core mass in our simulations.
Then, the Hill and Bondi radii for the planetary core at 5 au are
\begin{equation}
r_H=R\left(\frac{q}{3}\right)^{1/3}=0.0216 R=0.58 H\,,
\end{equation}
where $q=M_p/M_*$. With our fiducial disc temperature,
\begin{equation}
r_B=\frac{GM_p}{c_{s}^2}=q R\frac{R^2}{H^2}=0.0219 R=0.59 H\,.
\end{equation}
Both the Hill radius and the Bondi radius are about half of the disc scale height at 5 au. 
This is expected considering that the core mass of 10 $M_{\earth}$ is close to the disc thermal mass
\begin{equation}
M_{th}=\frac{c_s^3}{G\Omega}=\left(\frac{H}{R}\right)^3 M_{*}=17 M_{\earth}\,.
\end{equation}
The radii $r_B$ and $r_H$ are related to the thermal mass
with $r_B/H=M_p/M_{th}$, $r_H/H=3^{-1/3} (M_p/M_{th})^{1/3}$,
and $r_B/r_H=3^{1/3} (M_p/M_{th})^{2/3}$. Thus, with $M_p\sim M_{th}$, we have $r_B\sim r_H\sim H$.

Knowing disc and planet properties, we should be able to study how Jupiter gathers its envelope in 3-D simulations.
However, the atmosphere build-up timescale is the thermal (Kelvin-Helmholtz) timescale, which is millions of years, much longer than our 3-D simulations
can afford. Instead, we can only study snapshots during the atmosphere accretion, similar to the static model discussed in the introduction. 
Then, we can connect different snapshots using the energy conservation as in the quasi-static model. 

To simulate the envelope structure at one snapshot, we need to specify a luminosity throughout the envelope. 
The most important phase of Jupiter's atmosphere accretion is the phase when Jupiter is gathering its atmosphere mass to reach the crossover mass.
This is also the longest phase during the atmosphere accretion, which directly determines if Jupiter can go through the run-away accretion. 
The luminosity during this phase can be estimated with $L=GM_p\dot{M_p}/r_p$, which is 
$L=$5.92$\times 10^{27}$ ergs/s (or 1.54$\times 10^{-6}$ $L_{\odot}$) using $M_p=10 M_{\earth}$, $\dot{M_p}=10^{-5}M_{\earth}/yr$, and $r_p=2R_{\earth}$.
With this $\dot{M_p}$, 
a 10 $M_{\earth}$ atmosphere can be accreted onto the $10 M_{\earth}$ core during the 1 Myr disc lifetime, so that run-away accretion can occur before the gas disc disperses. In reality, this luminosity is at the high end of the envelope accretion luminosity. It is more representing the beginning of the atmosphere gathering stage or the run-away accretion stage (Section \ref{sec:jupiterevo}).  The actual accreted atmosphere is on top of the existing atmosphere so that not all the atmosphere is accreted to the core radius and we overestimate the energy released during the accretion. Furthermore, some protoplanetary discs' lifetimes can be 10 Myr so that  a 
lower accretion rate can still trigger run-away accretion. Thus, we have also carried out simulations with 10 times less luminosity, which is $L=$5.92$\times 10^{26}$ ergs/s (or 1.54$\times 10^{-7}$ $L_{\odot}$). We caution that a smaller luminosity leads to a colder and denser planetary atmosphere, so that, after the initialization of the simulation, the atmosphere needs to collapse further to reach a steady state and this collapse  takes a longer time. Within our simulation time ($\sim$10 orbits), the atmospheres in these low luminosity simulations are not fully settled to steady states, as discussed in section \ref{sec:3Dsphere}.

For our main simulations, we follow  \cite{Pollack1996} and \cite{Rafikov2006} to assume that the released energy is distributed throughout the envelope so that the luminosity actually follows (1-$r_p/r$)$L$.
By replacing $r_p$ with the inner boundary radius $r_{in}$ (in our simulations $r_{in}$ is $\sim$59 times $r_p=$2 $R_{\oplus}$, details in section 3), this luminosity profile is also more numerically friendly since there is no large intensity close to the inner
boundary. To achieve this luminosity, we add a heating rate to the whole envelope
\begin{equation}
\frac{dE}{dt}= \frac{L}{4\pi }\frac{r_{in}}{r^4}\,,
\end{equation}
where $E$ is the energy per unit volume.  The resulting luminosity increases sharply from the inner boundary, and, for the bulk of the envelope, the luminosity is a constant, similar
to the constant luminosity assumption in \cite{Piso2014}. For our test case in Section \ref{sec:test}, where we want to compare the simulation with the analytical solution, 
we indeed implement a constant flux inner boundary condition detailed in Appendix \ref{sec:constflux}.

After specifying the disc density, temperature, core mass, and core heating rate, we can simulate the planet's envelope in the disc.  The envelope reaches a steady
state with the given heating rate. We have tried 6 different setups (Table \ref{tab:setups}), including the fiducial case, the big dust case (thus lower opacity), the higher luminosity case, the hotter disc case, and 2 low luminosity cases that are identical to the fiducial case and the big dust case except for the low luminosity. These setups are summarized in Table \ref{tab:setups}.
The BigDust setup explores how dust growth affects the envelope structure. 
The HotDisk setup simulates the giant planet formation in a hotter disc environment.

As will be shown in section \ref{sec:results}, convection directly affects the envelope structure. For a spherical envelope, the maximum luminosity that can be carried out by radiation before the envelope becomes unstable to convection is
\begin{equation}
L_{max}=\frac{64\pi \sigma T^3}{3\kappa\rho(r)}\left(1-\frac{1}{\gamma}\right)\frac{\mu m_{u} Gm(r)}{k_B}\,,\label{eq:lmax}
\end{equation}
where m(r) is the mass within the radius r, $\rho(r)$ is the density at r, $\mu$ is the mean molecular weight, $m_u$ is the atomic mass constant, and $k_B$ is the Boltzmann constant. With T$\sim$100 K in the outer envelope, m$\sim$10$M_{\oplus}$,  $\kappa\sim$1 cm$^2$/g for our fiducial opacity, $\rho_{mid}=7\times10^{-11}$ g/cm$^3$, $\gamma=1.4$, and $\mu=2.35$, we can derive $L_{max}=1.7\times 10^{27}$ ergs/s. Since our fiducial luminosity is higher than this $L_{max}$, the envelope in our fiducial case is convectively unstable. However, the BigDust case with $\kappa\sim 0.1$  cm$^2$/g and the HotDisk case with $T_{irr}\sim$ 174 K have $\sim$10 times higher $L_{max}$ that are larger than our provided fiducial luminosity. Thus, envelopes in these two cases should be radiative. Our low luminosity cases should also be radiative accordingly. These estimates are consistent with our simulation results in Section \ref{sec:results}. Overall, by changing $L$, $\kappa$, and $T_{irr}$, we can control the strength of convection in the envelope.

\begin{figure*}
\includegraphics[trim=0mm 0mm 0mm 0mm, clip, width=3.4in]{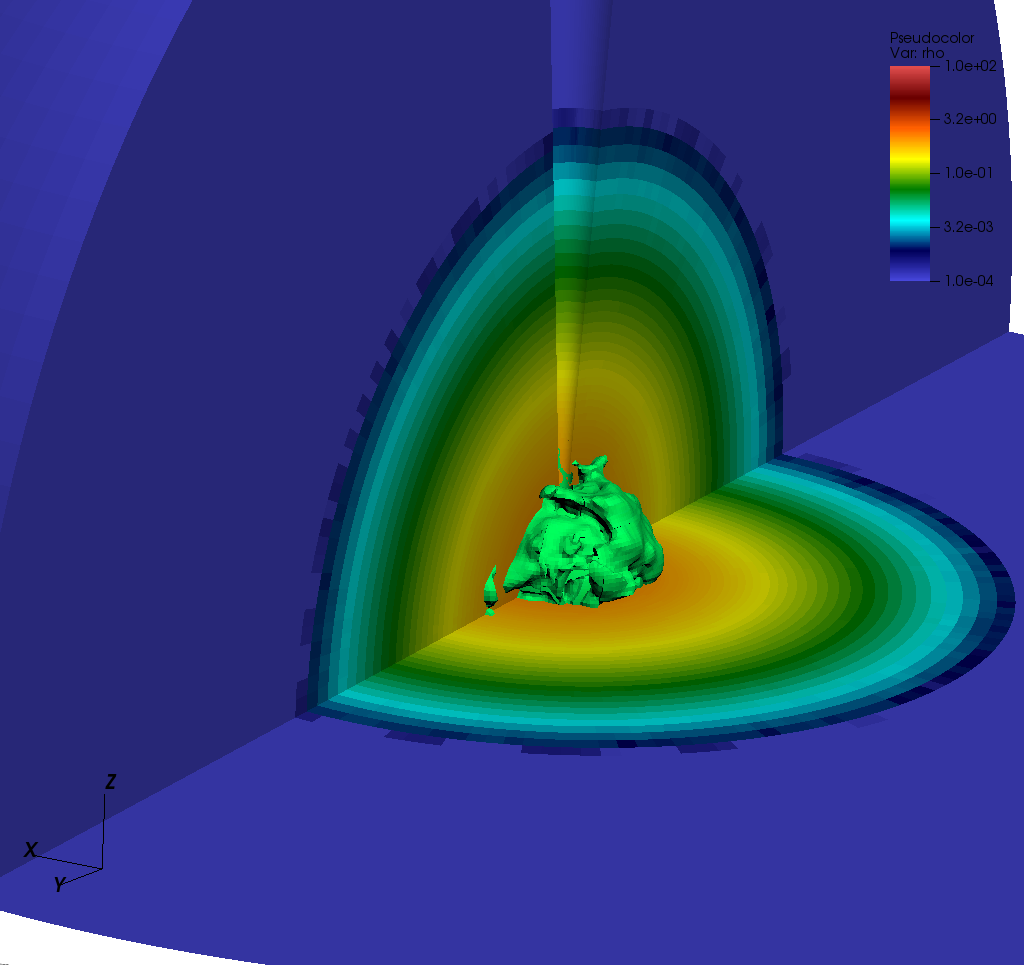}
\includegraphics[trim=0mm 0mm 0mm 0mm, clip, width=3.4in]{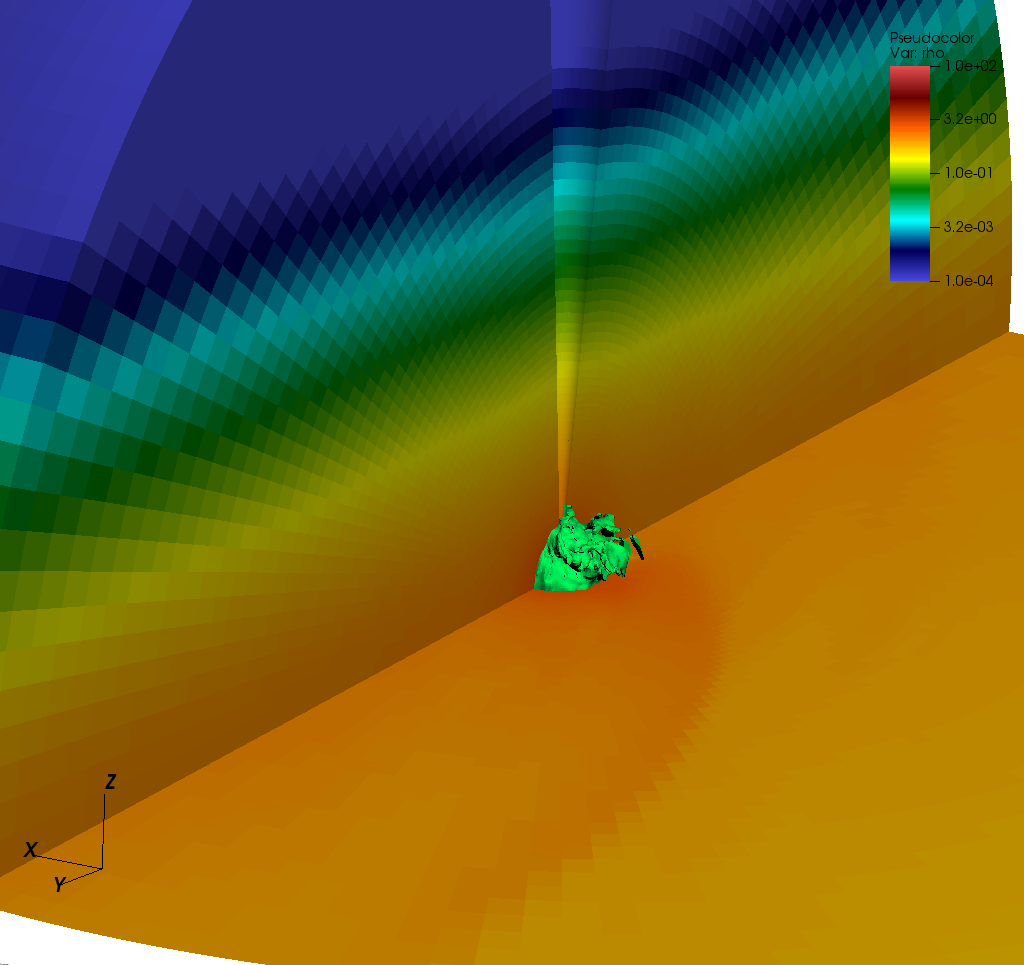}
\caption{ The density structure of the 3-D isolated sphere (the left panel) and 3-D envelope-in-disc (the right panel) simulations with our fiducial setup. The green surface at the center is the iso-surface with a constant radial radiation flux. The iso-surface is not smooth due to the convective motion within the envelope. }
\label{fig:render}
\end{figure*}

For each setup in Table \ref{tab:setups}, we carry out three simulations for comparison: the 1-D radiation simulation for the isolated and spherical envelope, the 3-D simulation for the isolated and spherical envelope
 (section \ref{sec:planetenvelope}), and the 3-D simulation for the planet envelope embedded in the disc (section \ref{sec:disksim}). We visualize these simulations in Figure \ref{fig:render}.

\subsection{1-D and 3-D Isolated Sphere Simulations}
\label{sec:planetenvelope}

\begin{table*}
\caption{Different setups for isolated-sphere and disc models}
\label{tab:setups}
\begin{tabular}{lcccccc}
\hline
                                                & Fiducial & HighL & BigDust & HotDisk  & LowL &LowLBD\\
                                                \hline
$L/(1.54\times10^{-6} L_{\odot})$&  1         & 10            & 1               &  1         & 0.1     & 0.1 \\
$a_{max}$ for dust                    &   1 mm & 1 mm    & 10 cm        &  1    mm   & 1 mm    & 10cm \\
T$_{irr}$                                  & 66 K     & 66 K      &  66 K          & 174 K  & 66 K  & 66 \\
t$_{end}$/t$_p$ for 3-D sphere             & 7.68          & 6.79               & 7.10                & 7.26  & 6.94  &5.47  \\ 
t$_{end}$/t$_p$ for 3-D disc             & 7.20          & 7.00               & 7.01                & 8.30    & 7.18  &8.00 \\ 
\hline
\end{tabular}
\end{table*}

For the isolated spherical envelope, both 1-D and 3-D radiation simulations have been carried out to study the effect of convection. Our 1-D radiation simulations have the radiative energy transport only. We do not add the energy transport due to convection from the sub-grid mixing-length theory. Thus, the 1-D and 3-D simulations have  identical setups, and
the only difference  is that, while 3-D simulations allow convection, 1-D simulations do not. If not specified, when we refer to 1-D simulations, we mean 1-D simulations with radiation only (no convection). Only in section \ref{sec:1dana} where we develop a simple model for reproducing 3-D simulation results, we consider convection in 1-D models.

The initial spherically symmetric envelope has a radial density profile of
\begin{equation}
\rho=\rho_{0} e^{-r^2/2H^2}\,,\label{eq:simpleatmosphere1}\\
\end{equation}
where $\rho_0=7\times10^{-11}$ g/cm$^3$, and H=0.185 au, representing the vertical density stratification in 
the protoplanetary disc at 5 au, following our choice of disc parameters after Equation \ref{eq:Tc}. 

To simulate the gravity from both the star and the planet, the envelope is subject to the gravitational force of
\begin{equation}
g(r)=\left(\frac{GM_p}{r^2}+r\Omega_{r=5 au}^2 \right) \times f_{s}\,,\label{eq:gr}
\end{equation}
where $\Omega_{r=5 au}$ is the angular frequency of Jupiter's orbit. The first term on the right is the gravity from the planet, while the second term is the force component towards the disc midplane due to the stellar gravity.
$f_s$ is a smoothing function,
\begin{equation}
f_s=\frac{(r-r_{in})^2}{(r-r_{in})^2+r_{s}^2}\,
\end{equation}
where $r_{in}$ is the inner boundary of the simulation domain, and $r_s$ is the gravitational smoothing length. With this smoothing function, the gravitational acceleration is zero at $r_{in}$, which allows the disc to maintain a better hydrostatic equilibrium close to the inner boundary \citep{Fung2019}.
Compared with previous 1-D envelope calculations which start with the disc midplane quantities,
our adopted gravitational force (Equation \ref{eq:gr}) guarantees a much smoother transition from the planet envelope to the background disc.
On the other hand, this setup is designed for simulating the vertical direction in the disc
(perpendicular to the disc midplane), and it does not include the anisotropic density stratification in the disc and the disc's Keplerian shear.
These shortcomings will be overcome by the disc setup in Section \ref{sec:disksim}.

The initial condition is not in hydrostatic equilibrium since Equation \ref{eq:simpleatmosphere1} does not take in account the gravity from the planet. After the planet gathers its envelope, the density structure beyond the planet's Hill radius should be smaller than the density from Equation \ref{eq:simpleatmosphere1}. However, in all our simulations, the planet's envelope mass within the Hill radius is at most 20\% of the total mass in the computational domain. Thus, the resulting density profile beyond the planet's Hill radius still follows Equation \ref{eq:simpleatmosphere1} very well.

The 10 $M_{\earth}$ core gathers
envelope material quickly and reaches steady state during the thermal time ($C_{v}nT/L$) where $C_v$ is the molar heat capacity and $L$ is the typical luminosity. 
With the 0.3 $M_{\earth}$ envelope (the typical 
envelope mass in our simulations) at our disc temperature and $L$ being close to our fiducial luminosity (at early times, $L$ is actually  higher than our fiducial luminosity), the thermal time is $\sim$ 5 times Jupiter's orbital period. Note that this thermal time in the simulation is significantly shorter than the Myr KH timescale of the planet's envelope since we only simulate the very outer region of the planet's envelope which has low mass and low temperature.
We run our simulations longer than the thermal time and ensure that the envelope has reached steady state (except for the low luminosity cases).

Our simulation domain covers r=0.001 to 0.3 $a_p$ with the Jupiter orbit's semi-major axis $a_p$=5 au with 140 grid cells that are uniformly spaced in logarithmic space. The domain range is equivalent to 0.027-8.1 H at 5 au.
The reflecting inner boundary and outflow (no inflow allowed) outer boundary conditions have been used in the radial direction. To guarantee that there is no mass flux
at the inner boundary, we set the mass flux from the Riemann solver to be zero at the inner boundary.
For 3-D simulations, we have 40 grid cells uniformly spaced from $\theta=0.1$
to $\pi/2$ in the poloidal direction, and 160 grid cells uniformly space from $\phi=0$
to $2 \pi$ in the azimuthal direction. We have adopted a reflecting boundary condition in the $\theta$ direction
and a periodic boundary condition in the $\phi$ direction. The heat capacity ratio $\gamma$ is chosen as 1.4 and the mean molecular weight is chosen as 2.35, 
considering that the gas mainly consists of molecular hydrogen at
our studied density and temperature range. A global density floor of 10$^{-10}$-10$^{-8}\rho_0$ has normally been adopted. 
The smoothing parameter $r_s$ for the gravitational force is chosen as 0.1 r$_{in}$. 

For radiative transfer calculations, we have used the angular discretization 
with respect to the local coordinate ($r$, $\theta$, $\phi$) in each cell. We have 8 longitudinal angles and 6 polar angles in the 3-D simulations (a total of 96 discrete directions including both inward and outward directions) and 20 polar angles in the 1-D simulations. 
At the outer radial boundary, we assume that there is incoming isotropic radiation from the background medium (e.g. stellar irradiation) with a temperature of 66 K. With such a boundary condition, our simulation domain
is kept at 66 K if there are no additional heating sources.
The tolerance error is set at $10^{-5}$ for the implicit method. 

\subsubsection{Test Problems}
\label{sec:test}
To explore the limitation of our grid setup and radiative transfer methods, we have done various slightly simplified 1-D spherical simulations with different resolutions and radiative transfer methods, and compare
them against the analytical results. 

When the envelope is radiative, we can calculate the envelope's thermal structure analytically. 
For a steady state, the second momentum equation of the radiative transfer equation becomes
\begin{equation}
\nabla\cdot{\bm P_{r}}=-\kappa_{T}\rho{\bm F_{r}}\,,
\end{equation} 
where ${\bm P_{r}}$ and ${\bm F_{r}}$ are the radiation pressure tensor and the radiation flux, and
$\kappa_{T}$ is the total opacity. With the Eddington approximation, we have
\begin{equation}
\frac{1}{3}\frac{\partial E_r}{\partial r}=-\kappa_{T}\rho\frac{L}{4\pi r^2c}\,.\label{eq:partialEr}
\end{equation}
where $\kappa_T$ is the total opacity (including the absorption and scattering opacity) and $c$ is the speed of light. 
With irradiation from a blackbody having a temperature of $T_{irr}$, we use the two stream approximation to
derive the temperature at the surface ($r=r_{out}$, $\tau=0$) 
\begin{equation}
\frac{L}{4\pi r_{out}^2}=F_{r}=\frac{ca_r}{\sqrt{3}}\left(T_{\tau=0}^4-T_{irr}^4\right)\,,
\end{equation}
where $a_r$ is the radiation density constant.
With $E_{r}(\tau=0)=a_rT_{\tau=0}^4$ as one boundary condition,
we can integrate Equation \ref{eq:partialEr} from the surface at $r_{out}$ to the interior at $r$ to derive
\begin{equation}
T(r)^4=\frac{3 L}{4\pi c a_r}\int_{r}^{r_{out}}\frac{\kappa_T\rho}{r^2}dr+\frac{\sqrt{3}L}{4\pi R^2 c a_r}+T_{irr}^4\,.\label{eq:Trtest}
\end{equation}
The choice of $r_{out}$ does not matter as long as the density drops sharply at large r so that the integral converges. We use our outer boundary 0.3 $a_p$ as $r_{out}$ in our calculation below.

\begin{figure*}
\includegraphics[trim=0mm 6mm 0mm 60mm, clip, width=7.in]{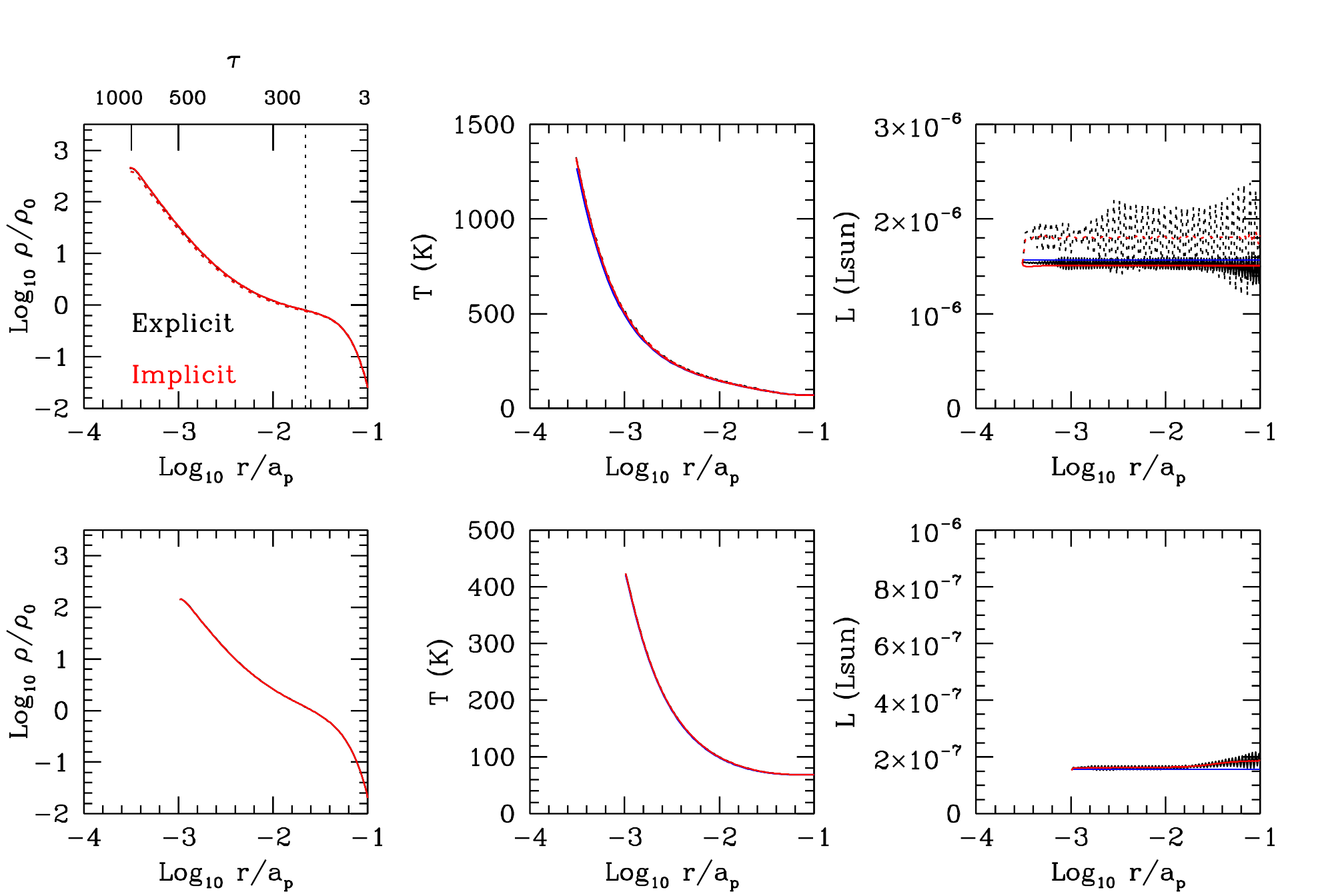}
\caption{The density, temperature, and luminosity in our 1-D envelope test problems with $r_{in}=0.0003\,a_p$ (upper panels) and $r_{in}=0.001\,a_p$ (lower panels).  
The black curves are from simulations with the explicit radiative transfer method,
while the red curves are from simulations with the implicit radiative transfer method.  The blue curves in the middle and right panels are analytical solutions.  
In the upper panels, the dotted and solid curves represent low and high resolution runs respectively. The vertical dashed line in the density panel labels $r_H$ (which is also very close to $r_B$ in this case).
}
\label{fig:test}
\end{figure*}

\begin{figure*}
\includegraphics[trim=0mm 0mm 0mm 0mm, clip, width=7.in]{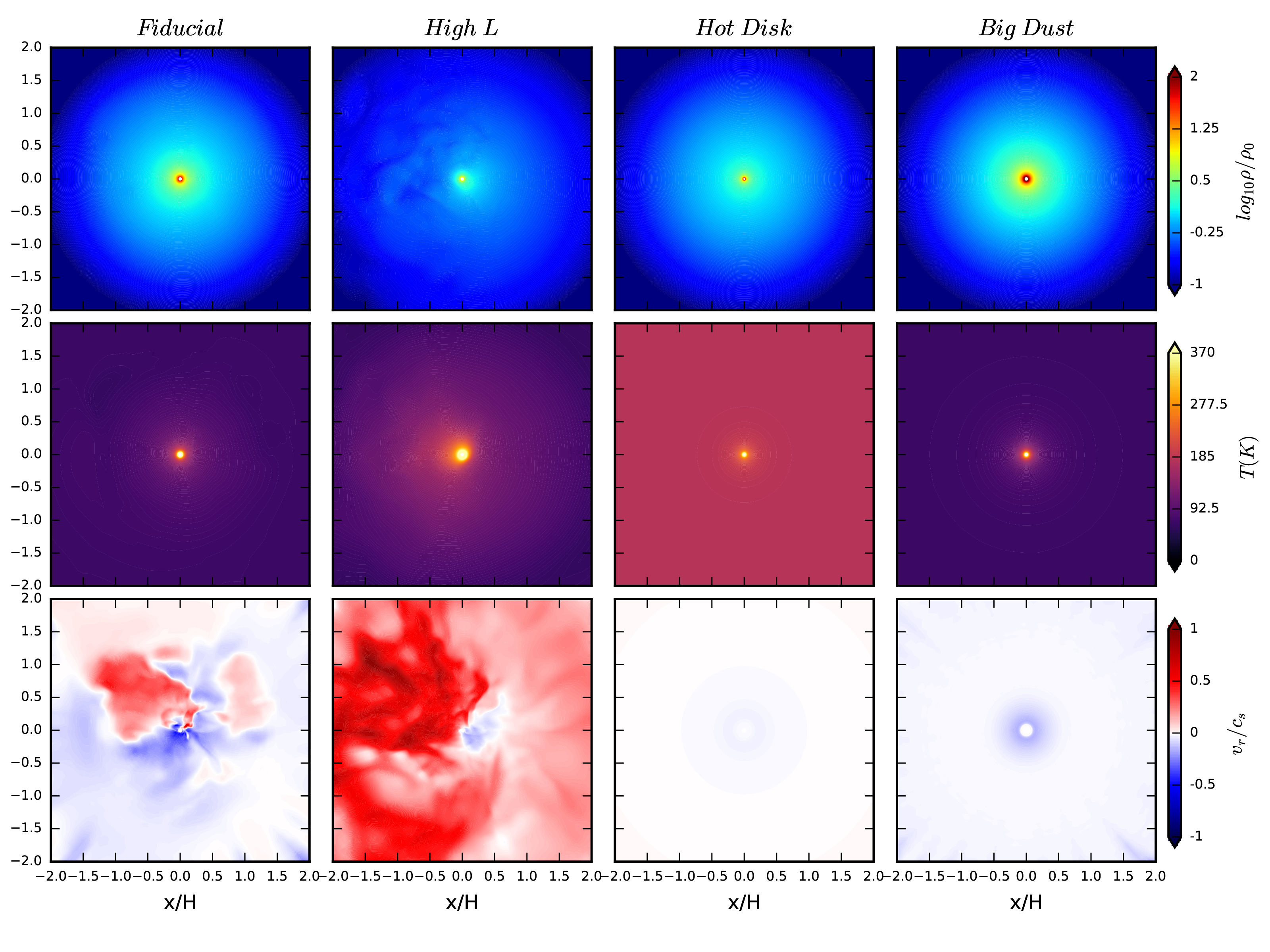}
\caption{ The density (upper panels), temperature (middle panels), and $v_r$ (normalized to the local sound speed, lower panels) 
for four 3-D isolated sphere models at the $\theta=\pi/2$ plane at the end of each simulation. Convection in the HighL
case is so strong that we adjust the $v_{r}/c_{s}$ colorbar for the HighL
case to [-2, 2]. 
The left two models show convection while the right two
models do not show convection.  }
\label{fig:sphere2D}
\end{figure*}

Figure \ref{fig:test} shows various 1-D test problems at t=50 planetary orbits. 
In the lower panels, we adopt our fiducial grid setup as in section \ref{sec:planetenvelope} but with 1/10th of our fiducial luminosity (thus 1.54$\times 10^{-7}$ $L_{\odot}$, the same as our low luminoisty cases).
We use the constant radiation flux boundary condition at the inner boundary (the implementation is detailed in Appendix \ref{sec:constflux}) so that we can compare with
the analytical solution  of Equation \ref{eq:Trtest}.  Both the Planck and Rosseland mean opacities are chosen as 1 cm$^2$/g for simplicity. All other parameters 
are the same as our fiducial cases in Section \ref{sec:planetenvelope}.
The blue curve in the middle panel is from the analytical solution,
which agrees well with our simulation results. 
After testing with various luminosities, we find that this luminosity is the lowest we can accurately simulate with our adopted grid setup. An even lower input
luminosity will lead to a sharp density/pressure gradient close to the core which will generate significant artificial heating that can propagate throughout
the whole envelope. 

To demonstrate the artificial heating due to insufficient resolution, we carry out test simulations with a smaller $r_{in}=0.0003 a_p$, as shown in the upper panels of Figure \ref{fig:test}. This leads to a sharper density/pressure gradient close to the core.  We adopt our fiducial luminosity in these simulations. We carry out simulations with two different resolutions: 
 100 (dotted curves) and 200 (solid curves) 
radial grid cells uniformly spaced in logarithmic space. In the very inner region, where the pressure gradient is the largest, 
the limited number of numerical grid cells means that it is harder to maintain a good hydrostatic equilibrium, and a noticeable amount of extra heat is generated. In the rightmost
panel, we can see that, when the resolution is low (100 grid cells), 
the real luminosity throughout the envelope is higher than our input luminosity from the inner boundary (the blue line).
When we double the resolution, we can recover our input luminosity. Thus, high resolution is needed to simulate the region with a 
steep pressure gradient, or a larger smoothing length is needed. The red curves are
from implicit methods, which show very similar results but with less variations. 

Fundamentally, such numerical heating is a manifestation of insufficient numerical resolution at sharp density transition close to the inner boundary. The numerical scheme
cannot maintain an infinitely sharp density transition. For example, we cannot simulate the whole atmosphere of a forming Jupiter if the atmosphere is isothermal, since
the density contrast from the core surface at $r_{core}$ to the Bondi radius is $exp(-(r_B-r_{core})/h_{core})$ where $(r_B-r_{core})/h_{core}=1.7\times10^6$ with the envelope scale height ($h_{core}$) calculated with our disc temperature.
Trying to simulate this structure with insufficient resolution leads to a much smoother density profile. For our radiative transfer calculations, a smoother
density profile close to the inner boundary corresponds to the density profile of a sphere with a higher luminosity. 
This higher luminosity can also be derived directly using the diffusion equation
where a lower $\rho$ leads to a higher luminosity with the same temperature structure. Thus, the smoother density profile manifests as numerical heating and higher luminosity. When we examine the velocity structure, we notice velocity fluctuations close to the inner boundary in the poorly balanced case, which drives the extra heating. 
It is difficult to give some specific rules on the numerical resolution needed. Different numerical schemes also provide different
results (in our setup the PPM reconstruction behaves better than the piecewise linear PLM reconstruction). We also find that a larger smoothing length of the planetary potential can significantly reduce the numerical heating (as in \citealt{Schulik2019}), but it also limits the ability to simulate the innermost high density region. 
Thus, numerical resolution tests are crucial. 
Overall, with our fiducial setup ($r_{in}=0.001\,a_p$), smoothing length ($r_s=0.1\, r_{in}$) and luminosity (L= 1.54$\times 10^{-7}$ or 1.54$\times 10^{-6}$ $L_{\odot}$), we have confirmed that the numerical heating rate is significantly lower than our input luminosity.

\subsection{3-D Disc Simulations}
\label{sec:disksim}
After simulating the isolated envelope assuming the spherical geometry, we begin to simulate the planet's envelope in a background protoplanetary disc.
Following the protoplanetary disc structure derived in section \ref{sec:diskstructure}, we set H/R=0.037 at 5 au and a flat temperature radial profile. The midplane
density at 5 au is $\rho_0=7\times10^{-11}$ g/cm$^3$, and it changes radially as $\rho_0 (R/5 {\rm au})^{-3}$ where $R$ is the distance to the star.
The disc vertical structure is self-consistently determined by the vertical hydrostatic equilibrium, and the azimuthal velocity around the central star
is also self-consistently determined by the radial force balance. Our simulation domain is centered around the planetary core at 5 au and is initiated with the background disc structure. Since the planetary core is orbiting around the central star, we need to consider non-inertial forces. The implementation of the non-inertial forces is given in Appendix \ref{sec:source} .

All parameters on the grid structure and boundary conditions are identical to our isolated sphere simulations in Section \ref{sec:planetenvelope}. The only differences
in the setup are the outer radial boundary condition for fluid quantities and the initial condition using the disc structure (Figure \ref{fig:render}). Considering that the outer radial boundary is only 0.3$\times$5 au away from the planet at 5 au,
the central star and most of the disc are outside the whole computation domain centered around the planet. Thus, we need to feed the Keplerian flow at our radial boundary to maintain
the disc structure. Thus, quantities at the outer boundary are always fixed with the disc's density, velocity, and energy as in the initial condition, which is similar to \cite{zhu2016}.

\section{Results}
\label{sec:results}
We first present our results from 3-D isolated sphere simulations (section \ref{sec:3Dsphere}), and then present results from 3-D disc simulations (section \ref{sec:3Ddisk}).
The 1-D spherical simulations will also be compared with these 3-D simulations. 

\subsection{3-D Isolated Sphere Simulations}
\label{sec:3Dsphere}

\begin{figure*}
\includegraphics[trim=0mm 4mm 0mm 0mm, clip, width=7.in]{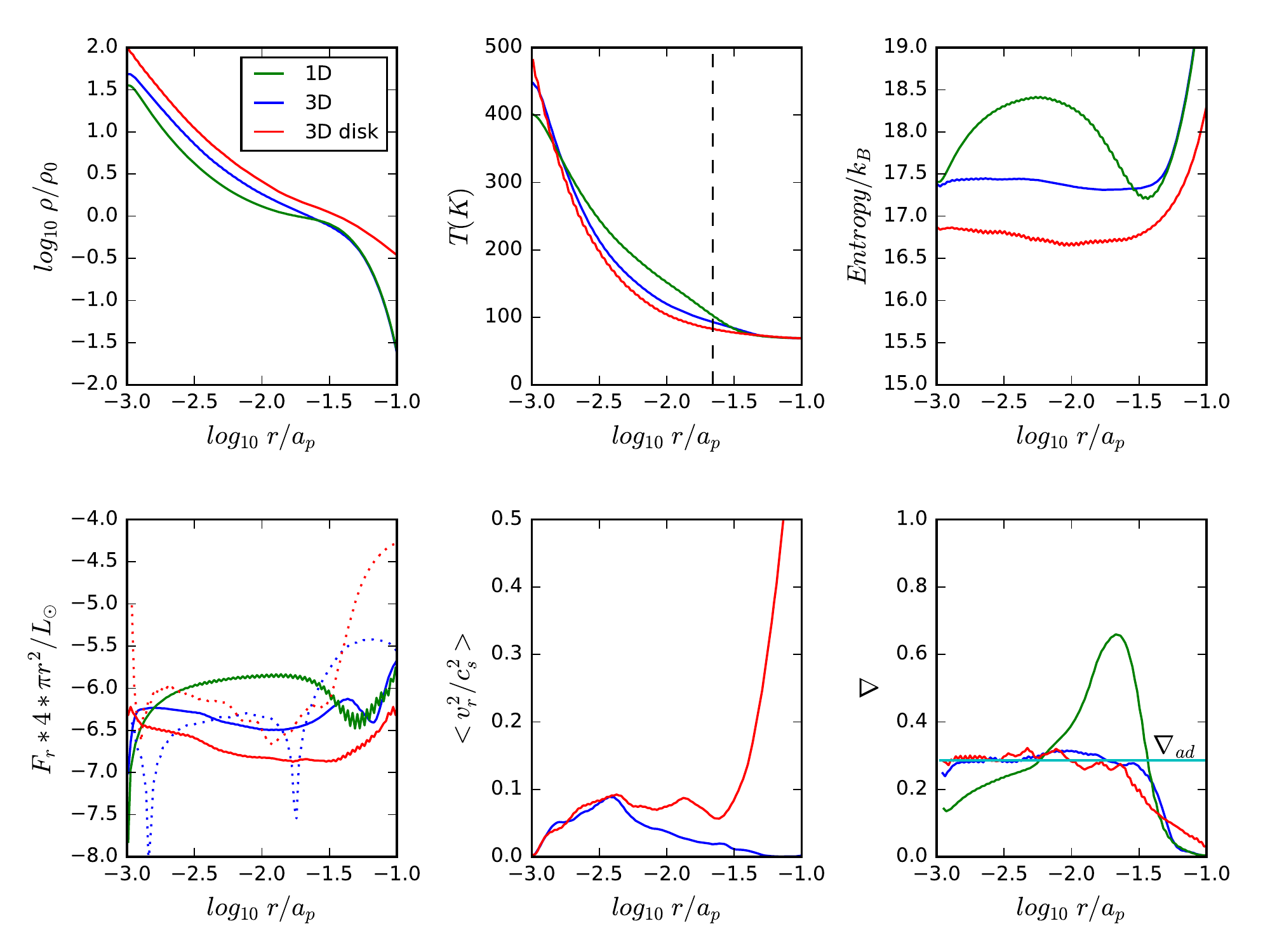}
\caption{ The radial profiles of various quantities for 1-D (green curves), 3-D isolated sphere (blue curves), and 3-D disc (red curves) models with our fiducial setup. 
All quantities are spherically averaged over grids, except for the lower left panel which shows the spherically integrated energy flux (solid curves: radiative fluxes,
dotted curves: convective fluxes from $E_{g}v_{r}$). The energy fluxes have
been averaged over 20 snapshots over a time span of 0.2 planetary orbits.  The energy fluxes, together with the pressure and temperature used in the lower right panel, have also been smoothed over 4 radial grids. The vertical dashed line in the temperature panel labels $r_{H}$ (which is also very close to $r_B$ in our fiducial case). Clearly, convection is prohibited in  the 1-D model so that $\nabla$ can be higher than $\nabla_{ad}$. 
 }
\label{fig:lowlum}
\end{figure*}

\begin{figure*}
\includegraphics[trim=0mm 3mm 0mm 0mm, clip, width=7.in]{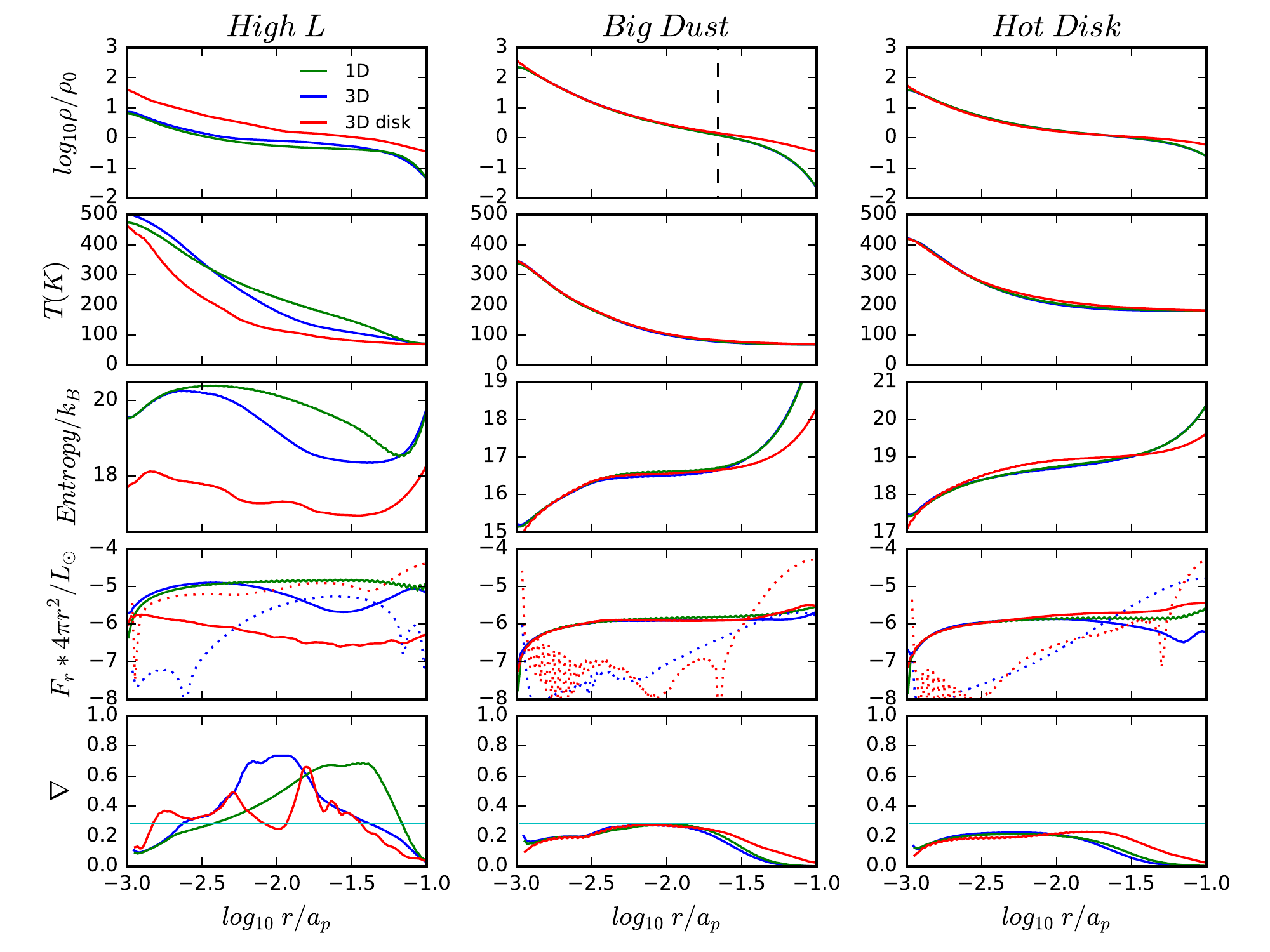}
\caption{ Similar to Figure \ref{fig:lowlum} but for three other setups.  }
\label{fig:multi}
\end{figure*}

\begin{figure}
\includegraphics[trim=5mm 3mm 67mm 0mm, clip, width=3.4in]{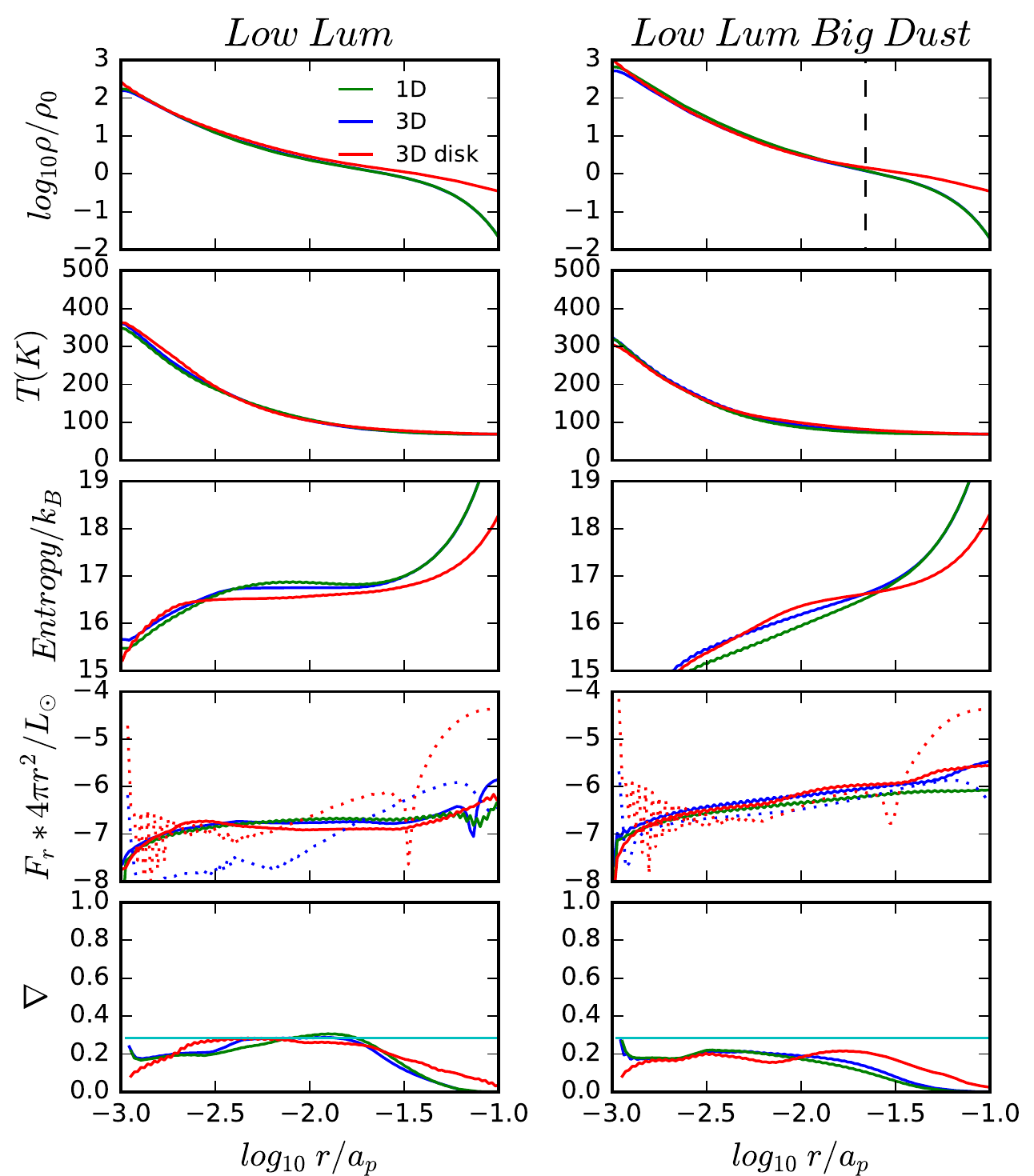}
\caption{ Similar to Figure \ref{fig:multi} but for two lower luminosity setups.  }
\label{fig:multilowlum}
\end{figure}

After running the 3-D isolated sphere simulations for several planetary orbits (Table \ref{tab:setups}), we present the envelope structure at $\theta=\pi/2$ plane in Figure \ref{fig:sphere2D}.
For our fiducial case, despite the fact that the density and temperature structures are very axisymmetric, there is strong subsonic convective motion in the envelope.
As shown in the leftmost panels of Figure \ref{fig:sphere2D}, no visible asymmetric structure can be seen in the density and temperature panels.
However, there is subsonic convective motion shown in the $v_r$ panel with the local mach number less than 0.5.

Convective motion is driven by the high luminosity in the envelope. To study the effects of convection on the envelope structure, we plot the spherically-averaged 1-D profiles of various quantities for both the 1-D and 3-D simulations in Figure \ref{fig:lowlum}. 
As shown in the $\nabla\equiv d {\rm ln} T/d {\rm ln} P$ panel, $\nabla$ in the 1-D simulation is actually higher than the adiabatic temperature gradient $\nabla_{ad}$ in order to transport the energy out by the radiative diffusion. The $\nabla_{ad}$ is  equal to $(\gamma-1)/\gamma$ for the ideal gas.
Based on the Schwarzschild criterion, whenever $\nabla$ is higher than $\nabla_{ad}$, the gas is convectively unstable, and, since convection is extremely efficient at transporting energy out, the envelope remains to be marginally convectively stable/unstable having $\nabla=\nabla_{ad}$. 
However, 1-D simulations do not allow convective motion, since one gas parcel that is under another gas parcel remains under in 1-D simulations. In 2-D or 3-D simulations, that gas parcel can move up through the second or third dimension. As shown in the $\nabla$ panel, allowing the convective motion, the 3-D spherical simulations exhibit the expected $\nabla\sim\nabla_{ad}$. This $\nabla\sim\nabla_{ad}$ holds all the way from the center to $log_{10}(r/5au)\sim -1.5$, indicating that
convection even reaches out beyond $r_B$ (labeled by the vertical dashed line). In this paper, we define entropy per particle as \begin{equation}
    Entropy=k_{B}ln\left(\frac{T^{\frac{\gamma}{\gamma-1}}}{P}{}\right)\,.
\end{equation}
The entropy is shown in the upper right panel of Figure \ref{fig:lowlum} and the curve is flat in the convective region, as expected.

Convection is carrying part of the energy out from the interior to the surface. The radiative flux ($F_{r}$) and convective energy flux ($\langle E_g v_{r}\rangle$ where $E_g$ is the internal energy density) are shown in the lower left panel of Figure \ref{fig:lowlum}. By comparing the 3-D isolated sphere simulation with the 1-D simulation, we can see that, in the 3-D isolated sphere simulation, almost half the energy is carried out by  convection and the other half by  radiative diffusion. If we add the radiative and convective flux together, we roughly recover the green curve within $r_B$. 

The presence of convection significantly changes the envelope's density and temperature structure. By comparing the 1-D and 3-D isolated sphere simulations,
we can see that convection in 3-D simulations makes the envelope colder and denser. By convectively transporting the energy out,
the disc does not need to be very hot to transport heat out radiatively.  The cooler envelope makes the envelope collapse more, leading to a higher
density. In this case, the whole envelope is convectively unstable in our simulation domain. 

After understanding the results for our fiducial case, we can study how different envelope parameters affect the envelope structure.
As shown in Figure \ref{fig:sphere2D}, the higher luminosity in ``HighL'' drives a much stronger convection. The colorbar for $v_{r}/c_{s,0}$ is adjusted
to [-2,2] accordingly, and we can clearly see that the turbulence becomes transonic. Such transonic motion does not provide enough time for the envelope to adjust itself from the turbulent motion. Thus,
the density and temperature structure is not axisymmetric. We can clearly see turbulent features in the $\rho$ and $T$ panels. 
In the left panels of Figure \ref{fig:multi}, we see similar behavior as the fiducial case. In the 1-D model, 
$\nabla$ is larger than $\nabla_{ad}$, while in the 3-D model convection tries to drive it back to $\nabla_{ad}$. We notice that the actual $\nabla$ is a little bit higher than $\nabla_{ad}$. We think this might be due to the fact that, with such a high luminosity, even  convection cannot transport energy out efficiently enough considering that the convective motion is already transonic. 
The convective envelope is again 
colder and denser in the 3-D model compared with the 1-D radiative model.

On the other hand, we can make the envelope less convective or even radiative by using either a hotter disc, a lower opacity, or a lower luminosity as discussed after Equation \ref{eq:lmax}. As shown in 
Figure \ref{fig:sphere2D}, the envelope structure is indeed axisymmetric and quiescent in the hotter disc or with the low opacity (from the bigger dust in the envelope). Most importantly,
due to the spherical symmetry and the lack of motion, the envelope structure is almost identical between 1-D and 3-D isolated sphere simulations. This is demonstrated
in the middle and right columns of Figure \ref{fig:multi}, where the green and blue curves perfectly overlap with each other in these radiative envelopes. The $\nabla$ is smaller
than $\nabla_{ad}$ everywhere for both the 1-D and 3-D models. The convective flux
is also significantly smaller than the radiative flux, again confirming that the envelope is convectively stable. Due to the radiative  
envelopes with $\nabla<\nabla_{ad}$, the entropy decreases towards the center (middle rows of Figure \ref{fig:multi}). 

Overall, the 1-D and 3-D isolated sphere models produce  identical envelope structure
for the radiative envelope, but they produce different structures for the convective envelope since  1-D simulations
do not allow convection. On the other hand, if we include the effects of convection (e.g. $\nabla=\nabla_{ad}$) in the 1-D semi-analytical model, we can reproduce the envelope structure in the 3-D spherical simulations using 1-D models, which will be studied in Section \ref{sec:1dana}.

We also plot the 1-D profiles of the lower luminosity cases in Figure \ref{fig:multilowlum}. The envelopes are mostly radiative. Even in 1-D radiative simulations, $\nabla$ is smaller than $\nabla_{ad}$ in most regions, except for a narrow $r$ range in the LowL case where $\nabla$ is barely higher than $\nabla_{ad}$. Since the envelopes are radiative, 1-D and 3-D isolated sphere simulations are almost identical. On the other hand, the envelopes have not fully reached steady states with such a low luminosity (especially for the LowLBD case). 
The envelopes are still contracting with significant $v_r$ so that the convective flux ($E_{g}v_{r}$) is still  high. The radiative flux in the LowLBD case is still higher than our input flux of 1.54$\times$10$^{-7} L_{\odot}$. The radiative flux is actually close to the convective flux suggesting that the energy increase due to the contraction is mostly radiated away. Longer 1-D radiative simulations suggest that steady states (when the contraction stops) can be reached at t$\sim$20 planetary orbits (20 $t_p$).

\subsection{3-D disc simulations}
\label{sec:3Ddisk}

\begin{figure*}
\includegraphics[trim=0mm 0mm 0mm 0mm, clip, width=7.in]{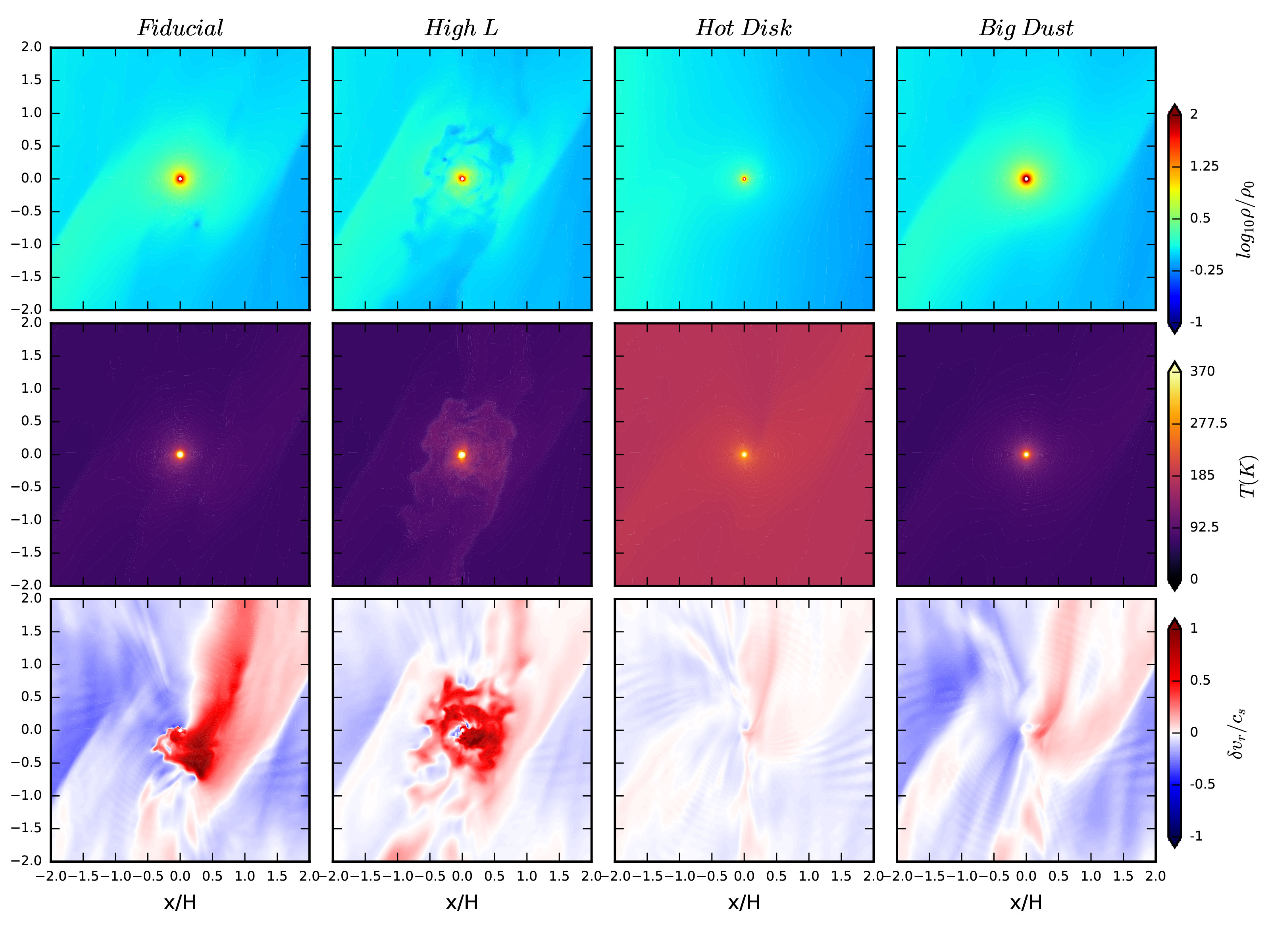}
\caption{ Similar to Figure \ref{fig:sphere2D} but for 3-D disc simulations. $\delta v_{r}$ is the velocity difference between this snapshot and the initial condition (from the Keplerian shear). Again, convection in the HighL
case is so strong that we adjust the $\delta v_{r}/c_{s}$ colorbar in the HighL case to [-2, 2]. }
\label{fig:disk2D}
\end{figure*}

Due to the Keplerian shear, the 3-D disc simulations exhibit spiral arms that extend from the planet. 
We can see density concentration along the spirals in the upper panels of Figure \ref{fig:disk2D}. 
These spirals can also perturb the disc's velocity structure \citep{Rabago2021}, as shown in the lower panels of Figure \ref{fig:disk2D} where the velocity perturbation ($\delta v_{r}$) suddenly changes sign along the spirals at y/H=$\pm$1 when x/H=$\pm$2.  
\cite{Goodman2001} pointed out that the strength of a spiral is mainly determined by the mass ratio between the planet mass and the disc thermal mass ($M_p/M_{th}$). 
For the Fiducial and the BigDust models, their disc temperatures are the same and thus $M_p/M_{th}$ is also the same. Then, their amplitudes of velocity perturbation at the spirals should also be similar, as shown in Figure \ref{fig:disk2D} (the leftmost and rightmost bottom panels).  For the HotDisk model, M$_{th}$, which is proportional to $c_s^3$, is a lot higher so that  $M_p/M_{th}$  is a lot smaller than our fiducial case. Thus, the excited spirals are much weaker, demonstrated as the less apparent spirals in the $\rho$ panel and smaller $\delta v_{r}/c_{s,0}$ in the velocity panel. Besides the spirals, we can also see the planet-induced horseshoe orbit at x/H$\sim\pm$0.5 when y/H=$\pm$2. 

On the other hand, the envelope structure is very similar between the 3-D isolated sphere and the 3-D disc models. 
For our fiducial case, the envelope is also convectively unstable and $\nabla\sim\nabla_{ad}$, as shown in the $\nabla$ panel of Figure \ref{fig:lowlum}.
For the convective envelope, the main difference between the 3-D isolated sphere and disc models is that the 
3-D disc model has a slightly colder and denser envelope. 
This is partly due to the disc geometry, which leads to a higher density after the spherical-averaging (e.g. at r/5au=0.1, the disc model has a higher density.). The higher averaged density with the same disc temperature leads to a lower entropy compared with the isolated sphere model at large $r$ (shown in the entropy panel). Since a convective envelope in 3-D tries to maintain
a constant entropy throughout the envelope (even beyond $r_B$), the envelope's entropy in a 3-D disc is lower than that in a 3-D sphere. Another factor, which may be even more important, is that the convective motion in the isolated sphere seems to extend to larger $r$ than that in the disc model. This can be seen in the  bottom left panel of Figure \ref{fig:sphere2D} where the convective motion extends to the scale of the disc scale height. The $\nabla$ panel in Figure \ref{fig:lowlum} also shows that the $\nabla\sim\nabla_{ad}$ region extends to a slightly larger $r$ in the 3-D isolated sphere. Thus, the 3-D isolated sphere maintains a high entropy from the surface. 
On the other hand, the envelope in a 3-D disc is colder with a lower entropy, which then leads to a higher density concentration.  
The higher density leads to a higher optical depth. The radiative diffusion equation suggests that the same temperature gradient leads to a smaller radiative energy flux in a more optical thick environment. Thus, compared with the 3-D isolated sphere model,  less energy in the disc model can be carried out by radiation and more energy
is now carried out by convection, as shown in the  lower left panel of Figure \ref{fig:lowlum}. The density and temperature differences between  isolated spheres and disc models can be accounted with our 1-D simple semi-analytical model in Section \ref{sec:1dana}.

For radiative envelopes, the envelope structure is almost identical among the 1-D/3-D isolated sphere and disc models, indicating that the disc geometry has
little effect on the radiative envelope structure. As shown in the right two columns of Figure \ref{fig:multi}, the red curves overlap the green
and blue curves perfectly within r$_B$. 
At a larger r, the disc geometry leads to a higher spherically averaged density which then causes a
lower entropy and a higher $\nabla$ there.  Unlike the convective envelope, this higher density beyond r$_B$ has little effect on the envelope structure.
This is because the structure of a radiative envelope is determined by the temperature instead of the entropy in the disc, and the disc has a quite uniform
temperature beyond $r_B$ where the stellar irradiation determines the temperature structure.
The photosphere of the envelope is quite high with our fiducial disc parameters. The photosphere is at $\sim$2.8 disc scale heights with our fiducial opacity, while the photosphere is at $\sim$2.1 disc scale heights with our big dust opacity.

\subsection{Recycling}
\label{sec:recycling}

\begin{figure*}
\includegraphics[trim=120mm 40mm 0mm 0mm, clip, width=7.in]{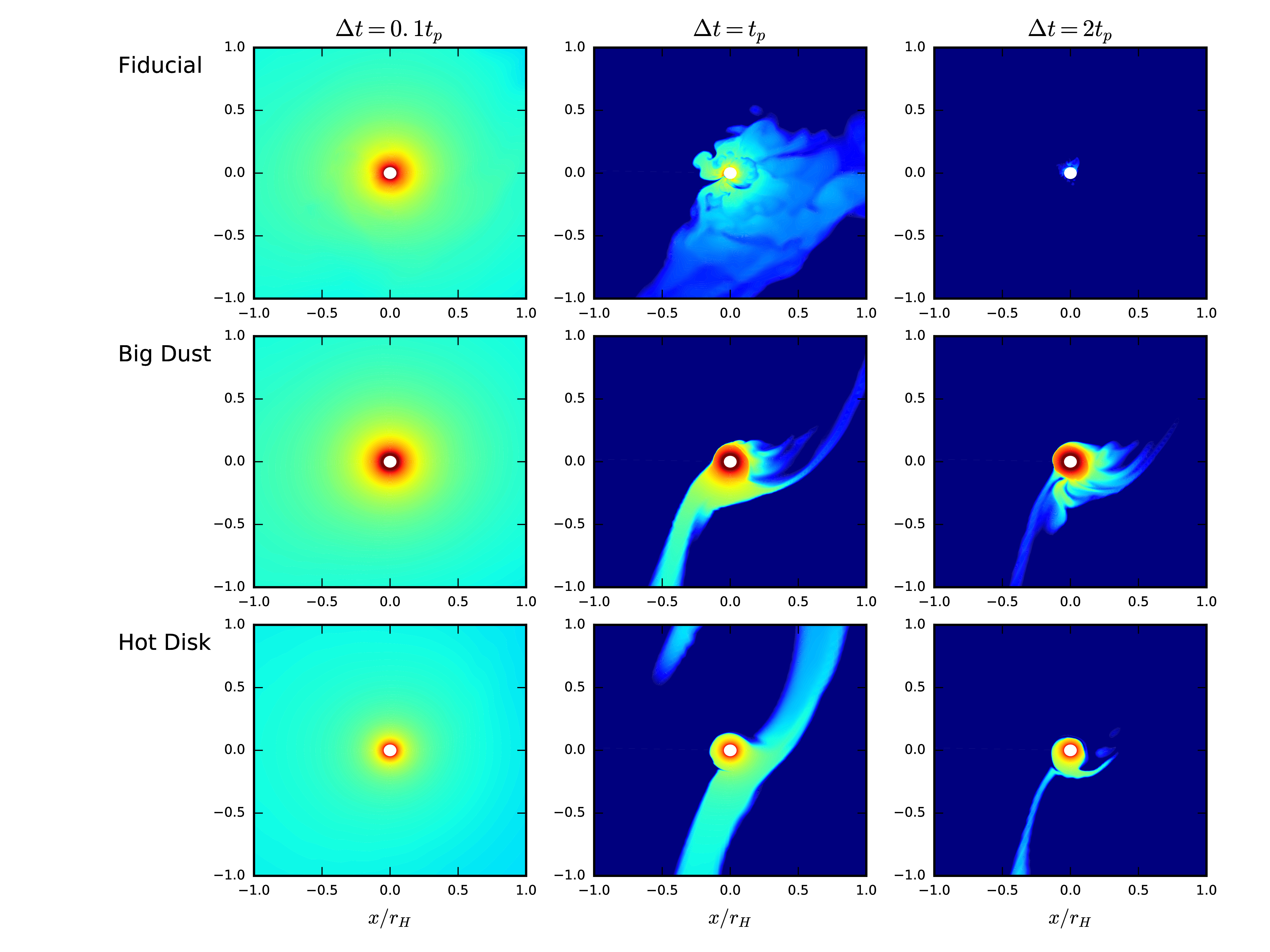}
\caption{ The evolution of the passive scalar at the disc midplane for the fiducial model (upper panels), the big dust model (middle panels), and the hot disc model (bottom panels) at different
times (left to right panels) after the passive scalar is injected.  }
\label{fig:concen}
\end{figure*}

\begin{figure*}
\includegraphics[trim=0mm 0mm 0mm 0mm, clip, width=7.in]{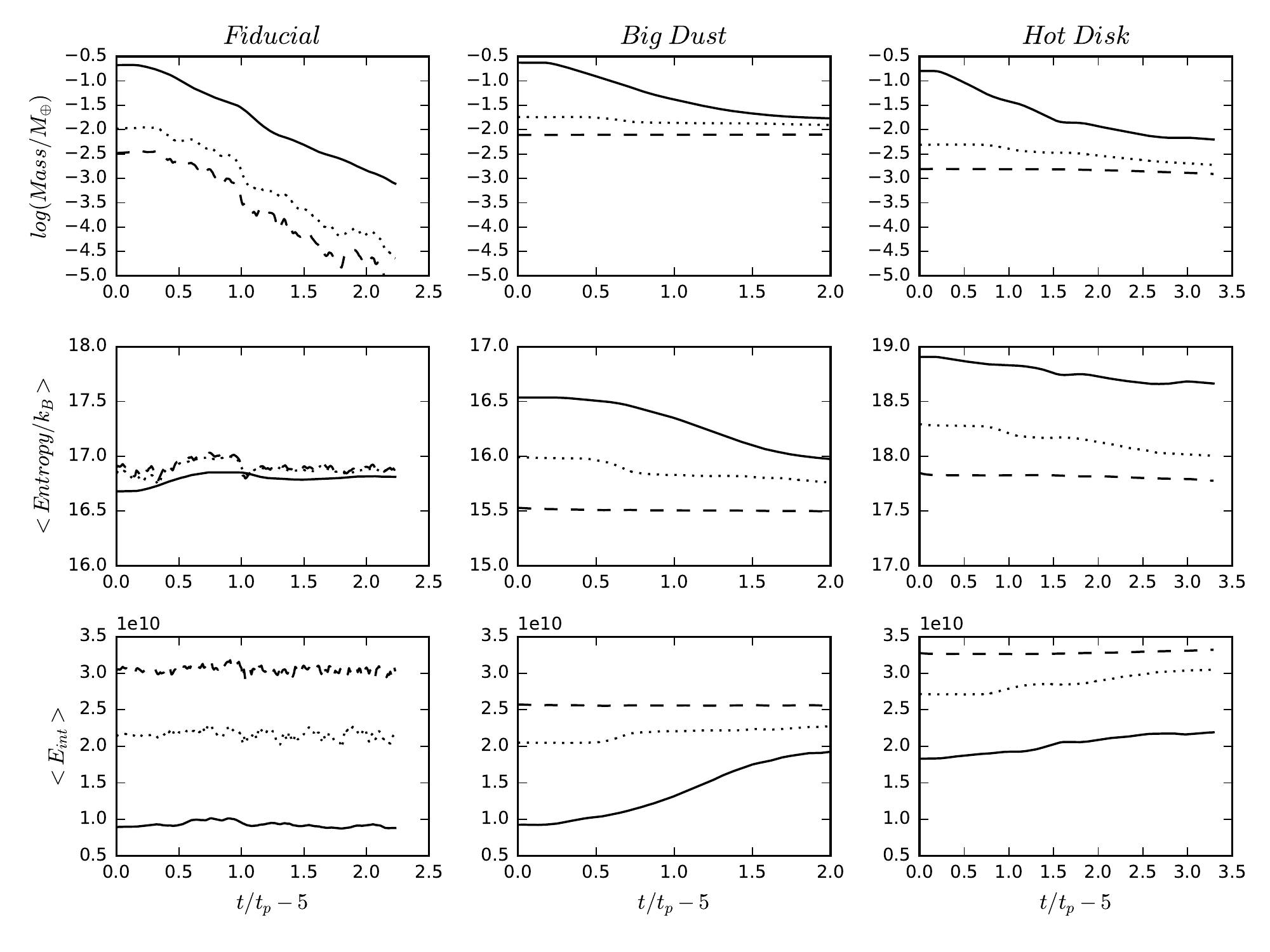}
\caption{ The time evolution of the passive scalar after being injected around the planet for three different disc models (left to right panels). 
 The upper, middle and lower panels show the integrated mass, the spherically averaged entropy, and the averaged  
internal energy per unit mass within a sphere of $r_B$ (solid curves), $r_B/5$ (dotted curves), and $r_B/10$ (dashed curves).}
\label{fig:totalmass}
\end{figure*}

\begin{figure*}
\includegraphics[trim=0mm 0mm 0mm 0mm, clip, width=7.in]{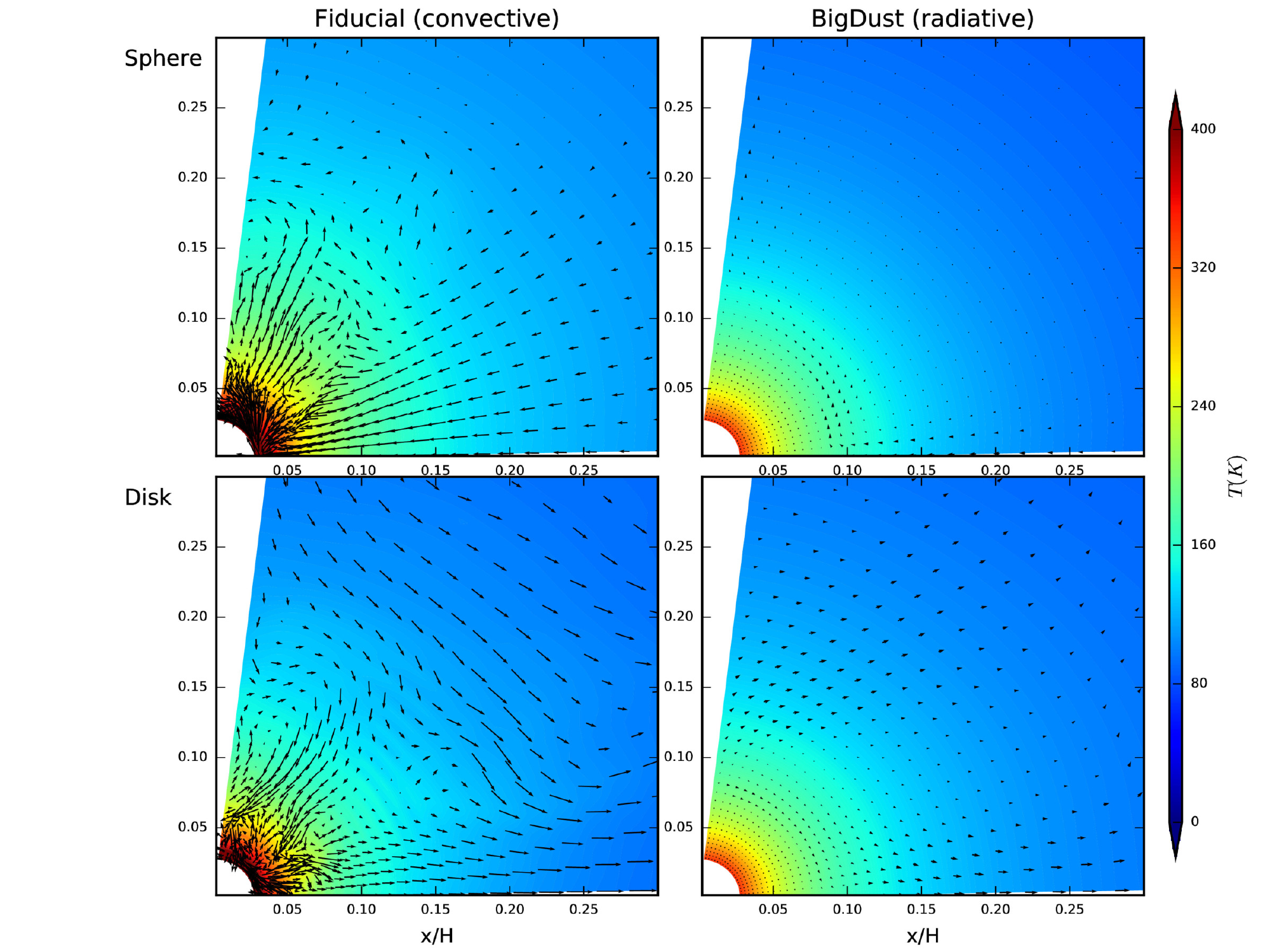}
\caption{ The vertical slices of the temperature structure  in 3-D isolated sphere models (upper panels) and 3-D disc models (lower panels) for 
the fiducial setup (left panels) and BigDust setup (right panels) in the direction away from the star. In the disc models, the slices are from the direction that is pointing towards to the central star. The lengths of the velocity vectors are on the same scale among all the four panels.  }
\label{fig:convection}
\end{figure*}

Such similarity between the isolated sphere and disc models for the radiative envelopes is quite surprising since we have observed significant net flow motion in  disc models, as shown in the bottom panels of
Figure \ref{fig:disk2D}. Both the spirals and the horseshoe orbits in disc models lead to non-zero $\delta v_r$. Moreover the disc's Keplerian shear itself carries material in the envelope away. 

To study how material within the envelope is recycled back into the disc, we add a passive scalar within $r/5au=0.05$ (roughly 2.5 r$_B$ for the fiducial model) at t=5 t$_p$ when the envelope has already settled to a steady state.  The passive scalar's initial density is the same as the envelope density at t=5 t$_p$. Then we continue the simulation for several orbits to see how the passive scalar is advected in the disc.  Since no passive scalar is added into the domain after the initialization at t=5 t$_p$, the passive scalar is carried away by the disc flow and diluted. By tracing the evolution of the passive scalar, we can understand
and quantify how mass is exchanged between the envelope region and the disc region. 

Figure \ref{fig:concen} shows the evolution of the passive scalar at the disc midplane for the convective envelope (fiducial model in the upper panels) and the radiative envelope (BigDust model in the middle panels and HotDisk model in the bottom panels) after the passive scalar is injected at t=5 $t_p$.  For the convective envelope, the convective motion quickly transports material near the center to larger radii where it is carried away by the Keplerian shear. Within 2 orbits, almost all of the original envelope material is recycled into the disc.  On the other hand,
for the radiative envelope, even though the outer envelope is recycled, the inner core (within 0.1-0.2 $r_H$) is protected from the recycling. We note that, the protected regions in the BigDust and the HotDisk models are both around 0.1-0.2 $r_H$,
although $r_H$ in the HotDisk model is $\sim 2.6 r_B$ while $r_H\sim r_B$ in the BigDust model.

To quantify the efficiency of the recycling, we integrate the mass, entropy, and internal energy of the passive scalar within a sphere of r$_{H}$, 0.2 r$_{H}$ and 0.1 r$_{H}$ for three different models, and plot these quantities with time in Figure \ref{fig:totalmass}. As shown in the upper left panel for the convective envelope, the total mass  of the passive scalar decreases exponentially no matter where the mass is calculated.  Within 2 planetary orbits, the mass has decreased by almost 2 orders of magnitude. The entropy and internal energy per unit mass is constant  with time, indicating that the passive scalar is well mixed in the envelope and the mass is depleted at the same rate at different envelope regions. For the radiative envelopes shown in the right two columns, the passive scalar's mass within r$_H$ decreases quickly, while the mass within 0.2 and 0.1 r$_H$ (especially within 0.1 r$_H$) is almost a constant. The entropy/internal energy per unit mass decreases/increases with time 
(especially within the r$_H$ sphere), indicating that the outer envelope where 
the entropy is the highest and the internal energy is the lowest has been recycled first. 

The recycling pattern revealed by the passive scalar can also be directly probed with the velocity vectors within the envelope. Figure \ref{fig:convection}
shows the velocity vectors for both isolated sphere and disc simulations. The vectors in all the panels are on the same scale. The left two panels show the convective envelopes, where we can clearly see the rolling motion at various scales. The right two panels show that the radiative envelopes have much slower motion.  Even though the lower right panel (the disc model) shows slightly faster motion than the upper right panel (the sphere model), the radial flow in the disc model decreases dramatically around the core at 0.07 H ($\sim$0.12 $r_H$). This is consistent with our passive scalar results that, for radiative envelopes, the region within 0.1-0.2 $r_H$ is protected against recycling. Although the global flow pattern is less robust in our local simulations which centered around the planet instead of the star, the global simulations from \cite{Fung2019} also reveal that the flow velocity decreases dramatically at similar scales.

The protected inner core could be qualitatively understood using the buoyancy force argument. When the gas parcel is moving towards the planet due to the planet's gravity (e.g. the Meridional flow), it is against the buoyancy force from a thermally stable atmosphere. Thus, it is slowed down and eventually stopped. Since the thermal effect plays a role here, we expect that the slowing down occurs when the entropy of the envelope is smaller than the disc entropy at the disc scale height. Figure \ref{fig:multi} shows that the entropy starts to decrease sharply at 0.2 $r_H$ for the radiative envelopes. Another way of thinking is that the infalling material stops at the position where its ramp pressure balances the thermal pressure. For a transonic inflow, this occurs when the envelope temperature becomes significantly higher than the disc temperature.

\section{Discussion}
\label{sec:discussion}

\begin{figure}
\includegraphics[trim=6mm 0mm 0mm 0mm, clip, width=3.5in]{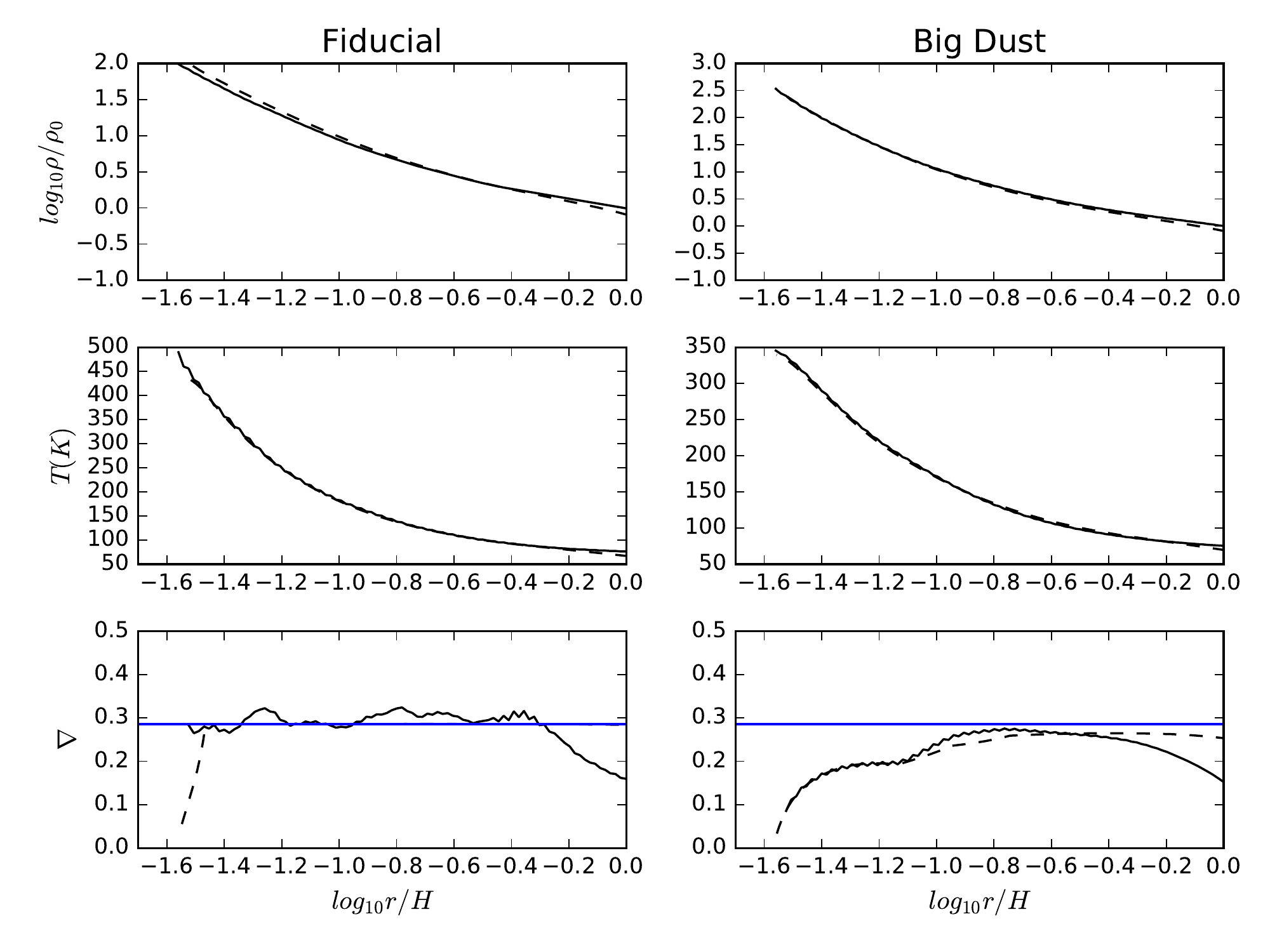}
\caption{ The envelopes' structure in 3-D simulations (solid curves) and 1-D semi-analytical models (dashed curves) of section \ref{sec:1dana}
for both the convective (left panels) and radiative (right panels) envelopes. The blue line in the $\nabla$ panel represents $\nabla_{ad}$. }
\label{fig:struc}
\end{figure}

\subsection{New procedures for 1-D semi-analytical models}
\label{sec:1dana}
Based on results from our direct 3-D simulations, we can improve previous 1-D models for calculating the envelope's structure, mainly regarding the outer
boundary condition in the 1-D models. Previous approaches use the minimum of the planet's Hill radius and Bondi radius as the outer boundary for 1-D calculations (e.g. \citealt{Bodenheimer1986}). 
The disc midplane density and temperature are used as the outer boundaries' density and temperature. Recently, a much smaller radius (e.g. 1/5-1/3 of the
Bondi radius) has been used as the outer boundary in 1-D calculations (e.g. \citealt{Lee2019, AliDib2020}) where the entropy is set to be the same as the disc's entropy. This is motivated by 3-D simulations in which material beyond that radius is recycled to the disc. For the fully convective envelope, the outer boundary position does not matter since the envelope's entropy is the same as the disc's entropy anyway. However, for radiative envelopes, our 3-D simulations do not support such a small radius approach at least for the planetary core at 5 au. 
For the radiative envelopes shown in Figure \ref{fig:multi}, $\nabla$ is smaller than $\nabla_{ad}$ throughout the whole region. There is no constant entropy region beyond 0.1-0.2 $r_B$. The density and temperature profiles are almost identical to those in the isolated sphere calculations. Recycling does happen (as in section \ref{sec:recycling}), but the timescale of recycling is longer than the thermal timescale of recycled material (Section \ref{sec:timescale}). Thus, the temperature structure is hardly affected by the recycling and is similar to that of an isolated sphere.

Considering that the planet's gravity does not abruptly stop at the Hill or Bondi radius, we choose a larger radius, the disc scale height, as the outer boundary condition. The spherically averaged density there is 
\begin{equation}
\rho=1.33 \rho_{mid}e^{-1/2}=0.81 \rho_{mid}\,,
\end{equation}
where the factor of 1.33 accounts for spherically averaging the flat disc density structure in every direction at r=H from the planet (e.g. in the midplane, the density is still the midplane density at r=H).  The temperature there is chosen as either the disc midplane temperature (if the envelope is convective at r=H)
or the temperature derived with Equation \ref{eq:Trtest} integrating from the disc surface to r=H (if the envelope is radiative at r=H). Finally, starting with these density and pressure at r=H, we integrate from r=H towards the core
with both the disc's and the planet's gravity (Equation \ref{eq:gr} with $f_s=1$) to derive the envelope structure. Similar to the traditional approach, $\nabla$ is calculated with the thermal diffusion equation, and we switch to $\nabla=\nabla_{ad}$ whenever $\nabla>\nabla_{ad}$. Compared with our 3-D radiation hydrodynamical simulations, this approach is significantly simpler and faster and we refer it as the 1-D semi-analytical model.

The resulting envelope structures using our fiducial opacity and $a_{max}=10$ cm opacity are shown in Figure \ref{fig:struc} as the dashed curves. We can see good agreement between this simple semi-analytical model and 3-D simulations for both the convective (left panels) and the radiative (right panels) cases. At the outer boundary, the density is slightly higher in 3-D simulations since even the disc region beyond r=H still feels the planet's gravity and concentrates slightly.

\subsection{Various timescales}
\label{sec:timescale}

\begin{figure}
\includegraphics[trim=0mm 0mm 0mm 0mm, clip, width=3.5in]{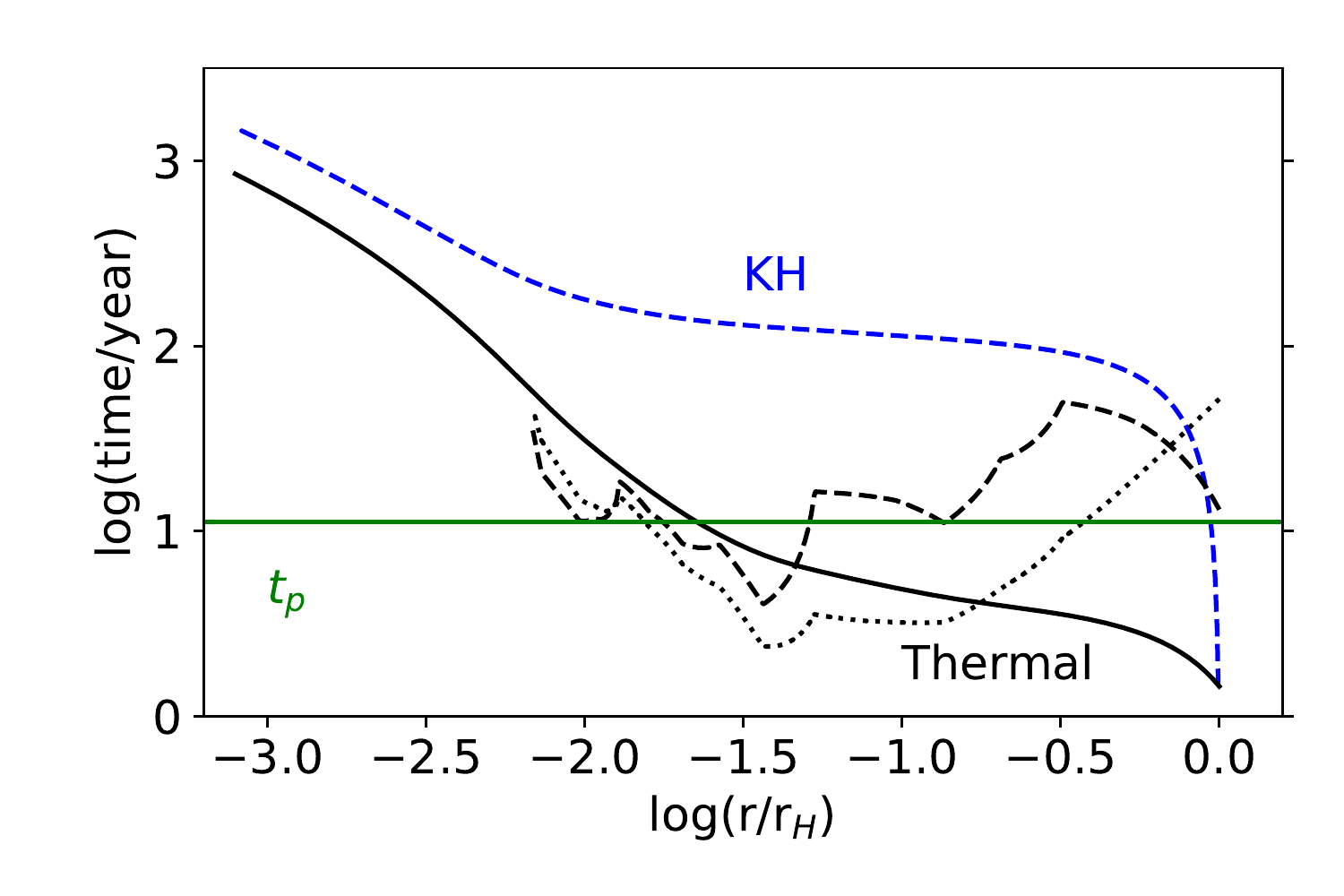}
\caption{ The Kelvin-Helmholtz timescale (dashed blue curves), and the thermal timescale for the recycled material (black curves) based on the 1-D  model with the big dust opacity. The horizontal dashed line represent the Jupiter's orbital time. The solid, dotted, and dashed black curves are different estimates for the thermal timescale using different approximations (detailed in the text). }
\label{fig:timescale}
\end{figure}

The planet's atmosphere/envelope accretion involves dramatically different spatial scales (e.g. core and disc scales) and time scales (e.g. recycling and KH timescales). The structure of the envelope largely depends on the relative amplitudes among these scales. Unfortunately, direct numerical simulations can only simulate very limited spatial and time scales. Thus, it is important to estimate the various time scales for the entire envelope. Particularly, we want to compare the recycled material's cooling timescale with the recycling timescale to understand if the recycling process can affect the envelope's thermal structure.

Cooling timescales are the ratio of energy content to an energy loss rate.  The cooling time for an entire bound object is a KH timescale, $E/L$, where due to the virial theorem it is most correct to use total energy, but (also by the virial theorem) either the thermal or gravitational energy is usually a good approximation.  To estimate the cooling timescale, we use our 1-D semi-analytical model in section \ref{sec:1dana} to calculate the planet's envelope structure all the way to the core (2 earth radii) with either the fiducial opacity or the big dust opacity. All other parameters, including the disc condition, the luminosity, and the adiabatic index, are the same as in our fiducial case. From the resulting structure, we integrate the thermal energy from the Hill radius inwards
\begin{equation}
E_{tot}(r)=\int_{r_H}^r E(r')4\pi r'^2 dr'
\end{equation}
where and $E(r')$ is the energy per unit volume at $r'$.  Then, we use $E_{tot}(r)/L$ to derive the KH timescale at different $r$ in the envelope. It represents
the timescale for the given luminosity to affect the thermal energy of the envelope material above $r$. This timescale for the big opacity case is given in Figure \ref{fig:timescale} as the blue dashed curve. This is the longest timescale in the process of envelope accretion. We note that our KH timescale at the core is only 10$^3$ years, significantly shorter than the Myr KH timescale normally assumed in the Jupiter's atmosphere accretion. This is mainly due to our assumed high luminosity. As discussed in section \ref{sec:diskstructure}, our fiducial luminosity is $\sim$ 10 times higher than the typical envelope's luminosity during the contraction. This high luminosity not only transports the energy out quicker, but also leads to a hotter and less massive envelope. Both effects decrease the KH timescale in our models. On the other hand, this KH timescale is the whole envelope's cooling and contraction timescale, which is not the recycled material's cooling timescale. 

Cooling times for the recycled material can be derived using the cooling times for local thermal perturbations.  We recap this timescale for the case of optically thick perturbations larger than $\ell_{\lambda} = (\kappa \rho)^{-1}$.  For a thermal perturbaton $\delta T$ on lengthscale $\ell_x$ in a plane-parallel atmosphere, the excess energy per volume is $\rho c_V \delta T$ with $c_V$ the specfic heat per unit mass.  The energy loss per volume is  $\nabla \cdot F \sim \delta F/\ell_x$. The radiative flux is $F = k_{\rm rad} \nabla T$ with $k_{\rm rad} = 16 \sigma T^3/(3 \kappa \rho)$.  Thus $\delta F \sim k_{\rm rad} \delta T / \ell_x$ and 
\begin{align}\label{eq:tcool}
t_{\rm cool} &\sim \frac{3 c_V \rho^2 \kappa }{16 \sigma T^3} \ell_x^2
\end{align} 
in agreement with standard expressions (in the diffusive regime). We can use this cooling timescale to describe cooling during the ``atmospheric recycling'' process. 

``Atmospheric recycling", i.e. the flow of disc gas inside the envelope of a protoplanet (i.e. inside its Bondi radius) can limit the cooling of a protoplanet by constantly delivering disc material which has a higher entropy than any parts of the envelope that have already cooled. The existence of the radiative envelopes in some simulations reveal that in regions with recycling the envelope still has a stably stratified, radiative structure with $\nabla < \nabla_{\rm ad}$. Furthermore, $\nabla$ in these 3-D radiative envelopes which are subject to recycling is almost identical to the $\nabla$ in corresponding 1-D radiative simulations, indicating that the recycled high entropy disk flows cool efficiently on their recycling timescale.  To quantify this we define the extra  energy content that must be lost if the actual temperature $T$ is below the adiabatic  $T_{\rm ad}$ that has the disc entropy (at that radius and pressure).
\begin{align}
E_{\rm extra} &= 4 \pi \int_r^{r_{\rm out}} \rho c_V (T_{\rm ad} - T)r'^2 dr' \label{eq:EZhu}\\
& \simeq 4 \pi \int_r^{r_{\rm out}} \rho c_V \left[\int_{r'}^{r_{\rm out}}(\nabla_{\rm ad} - \nabla) \frac{\mu g }{\mathcal{R}} dr'' \right]r'^2 dr'
\end{align} 
where we use $T = T_o(r_{\rm out}) - \int dT/dr dr = T_o + \int\nabla \mu g/\mathcal{R}dr$, with an ideal gas law $P = \rho \mathcal{R} T/ \mu$, the hydrostatic equilibrium $dP/dr=\rho g$, and similarly for $T_{\rm ad}$.   We further approximate the integral as 
\begin{align}\label{eq:extraapp}
E_{\rm extra} &\sim 4\pi c_V \rho T
r^3 (\nabla_{\rm ad} - \nabla)
\end{align} 
which could be too severe an approximation. It assumes that $\int_r g dr \sim G M_{\rm p}/r$ and that the density integral is dominated by the deepest $H_{\rm p} = P/(\rho g)$.  

The available luminosity to carry away this $E_{\rm extra}$ is likely not the entire luminosity as usual sources of luminosity (KH contraction and planetesimal heating) are not responsible for this additional cooling of the recycling flows.  Moreover there is likely no additional luminosity to easily measure in a numerical model, if it is in a quasi-steady state. Given these issues, the relevant luminosity is best defined relative to $E_{\rm extra}$ as the additional luminosity that would be generated by an adiabatic envelope, i.e.
\begin{align}
L_{\rm extra} &= \frac{64 \pi \sigma T^4 G m(r)}{3 \kappa P}\left(\nabla_{\rm ad} - \nabla\right)  \label{eq:Lextra}\\
&= \frac{64 \pi \sigma T^4 r^2}{3 \tau_\parallel }\left(\nabla_{\rm ad} - \nabla\right)
\end{align} 
where the final form uses the optical depth $\tau_\parallel = \kappa P/g$ that would apply in a plane-parallel atmosphere.

The resulting cooling time for recycling flows is then
\begin{align}
t_{\rm crc} &= \frac{E_{\rm extra}}{L_{\rm extra}} \sim \frac{15 P r \tau_\parallel}{32 \sigma T^4} \label{eq:tcrcsimple}
\end{align} 
where the approximate expression uses Equation \ref{eq:extraapp} instead of Equation \ref{eq:EZhu} and the diatomic $c_V = 5/2 \mathcal{R}/\mu$. If we compare to the standard Equation \ref{eq:tcool} then
\begin{align}
\frac{t_{\rm crc}}{t_{\rm cool}} &= \frac{\tau_\parallel r }{\tau_x \ell_x} 
\end{align} 
where $\tau_x = \rho \kappa \ell_x$ for the perturbation, so the only difference is in effective optical depths and length scales.  

Thus, we can estimate the cooling time for recycling flows using our derived 1-D envelope structure. Figure \ref{fig:timescale} shows these time estimates for the big dust opacity case. The solid black curve represents $E_{\rm extra}/L$ where $E_{\rm extra}$ is from Equation \ref{eq:EZhu} and L is the constant luminosity throughout the envelope. The dashed black curve uses the same $E_{\rm extra}$ but $L$ is from Equation \ref{eq:Lextra} at different r. The dotted black curve is from Equation \ref{eq:tcrcsimple}. Overall, the cooling timescale is comparable or shorter than the orbital timescale (which is also the recycling timescale) at $r>0.01 r_H$. Thus, recycling has limited impact on the thermal structure of the envelope. 

At the same time, the core and envelope can be convective (e.g. our fiducial opacity case). When the envelope is convective, the energy transport timescale is the timescale of the convective motion. The convection speed can be estimated using the Mixing-length-theory (MLT). However, the MLT  requires us to solve the deviation of the envelope's temperature gradient from the adiabatic temperature gradient, which is beyond our simple 1-D model. Thus, we roughly estimate the convection velocity using the empirical relationship \citep{Porter2000,Jones2017,Fuller2017}
\begin{equation}
v_{conv}(r)=\left(\frac{L}{4\pi r^2 \rho}\right)^{1/3}\,.\label{eq:convspeed}
\end{equation}  
Using the 1-D model with our fiducial setup, we can estimate $v_{conv}(r)$.
It turns out that  $v_{conv}(r)$ is roughly a constant throughout the envelope with the value of $\sim$0.1 $c_{s,0}$ where $c_{s,0}$ is the sound speed at the disc temperature. This is consistent with our simulations (Figure \ref{fig:sphere2D}). Thus, the convection timescale ($r_H$/(0.1 $c_{s,0}$)$\sim 6/\Omega$) is the orbital timescale.

By comparing these timescales, we can discuss the effect of recycling which occurs on the orbital timescale (section \ref{sec:recycling}). For a convective envelope, the recycling timescale is  similar to the convective timescale. Thus, recycling can quickly exchange the envelope material with the disc material even to the deepest part of the outermost convective envelope. In our fiducial 3-D disc simulation, the convective envelope extends to our inner boundary, which explains why all the envelope material has been recycled efficiently. In our 1-D semi-analytical model, there is a radiative zone from 0.005 $r_H$ to 0.05 $r_H$ separating the inner and outer convective regions, so that the recycling may only be effective for the outer
convective zone (>0.05 $r_H$).

For a radiative envelope, the mass recycling is only efficient beyond 0.1 $r_H$. Even so, the thermal structure of
that outer region is hardly affected by the recycling process, implying that the thermal timescale to establish the envelope's thermal structure is shorter than the recycling timescale. Quantifying this thermal timescale is difficult. Our crude estimate (black curves in Figure \ref{fig:timescale}) indeed suggests that the cooling timescale is shorter/comparable to the recycling timescale. Overall, for the planet at 5 au, the recycling has large effects on the mass exchange for the outer envelope, but it has limited effects on the thermal structure of the envelope. At 0.1 au where the orbital timescale decreases dramatically, the recycling may have stronger effects on the envelope's thermal structure.

\subsection{Jupiter's Evolution}
\label{sec:jupiterevo}

\begin{figure}
\includegraphics[trim=0mm 0mm 0mm 0mm, clip, width=3.5in]{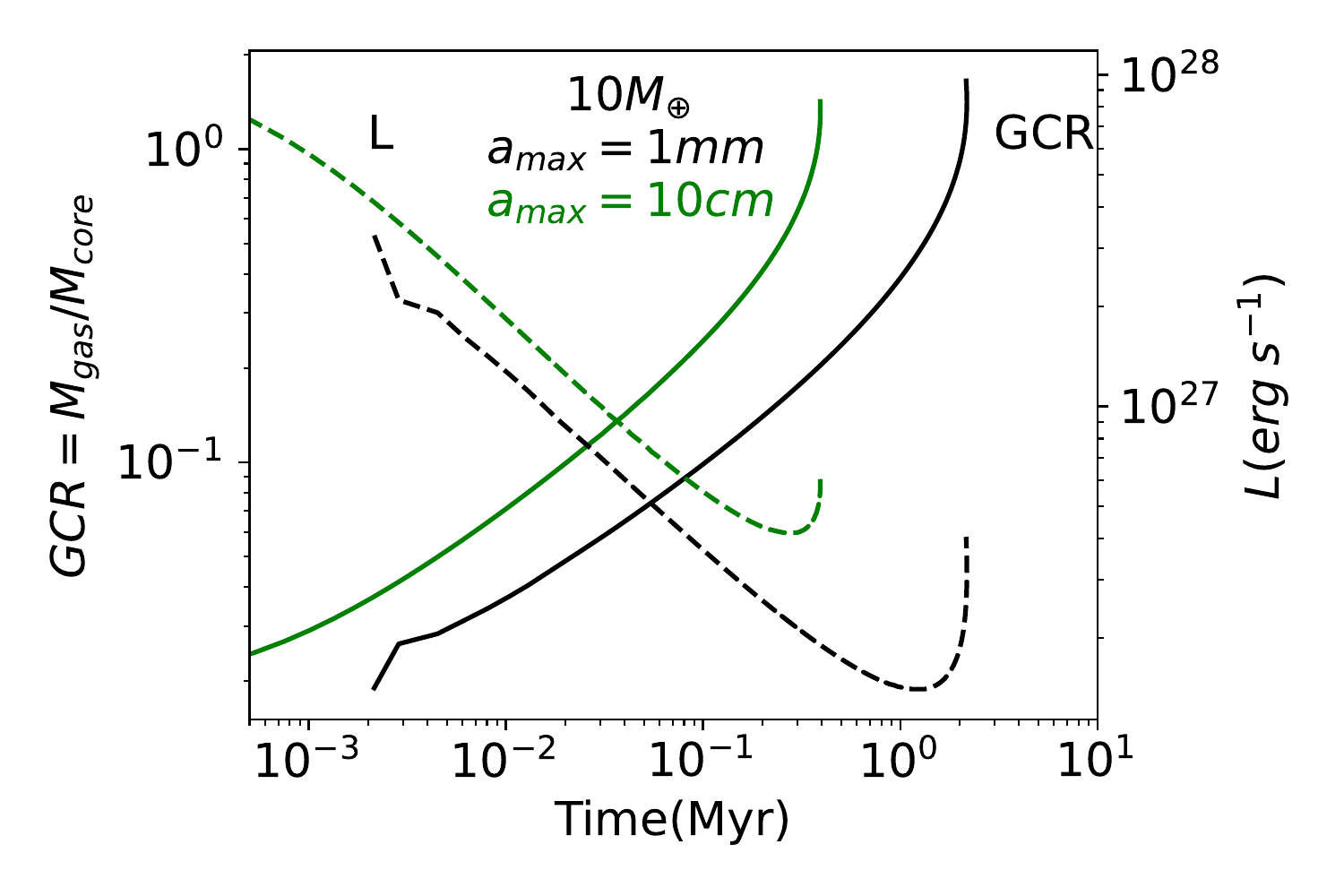}
\caption{  The gas-to-core mass ratio (solid curves) and the luminosity (dashed curves) for our Jupiter evolutionary model with the fiducial opacity (black curves) and the lower opacity (green curves).}
\label{fig:evolution}
\end{figure}

After calculating the envelope structure at different snapshots, we can use the energy conservation (e.g. \citealt{Piso2014}) to connect them and derive the time evolution of the Jupiter's envelope accretion. In detail, given an envelope mass, we search for the luminosity which can sustain this envelope, and derive the resulting envelope structure.
Then, we calculate the energy
difference between two snapshots having slightly different envelope masses, and divide this difference with the luminosity to calculate
the time span between these two snapshots. Finally, connecting all the snapshots
from small to large envelope masses, we can derive the time evolution of envelope accretion. Runaway accretion starts when the luminosity begins to increase
with the increasing envelope mass, which normally occurs when the atmosphere mass roughly equals the core mass (the atmosphere reaches the cross-over mass). 
We have ignored the effects of mass and volume change on the energy equation, as in \cite{Lee2014}. We use the new 1-D models in section \ref{sec:1dana} to calculate the envelope structure, but  now with the full EOS from \cite{Piso2014} and the new opacity in Section \ref{sec:opacity}. When we calculate the envelope's energy, we only include the envelope within 1/10 th of the Hill radius since only this region is bound in our radiative envelopes. Figure \ref{fig:evolution} shows the time evolution of
the envelope-to-core mass ratio and the envelope luminosity for a 10 earth mass core at 5 au. The envelope mass only includes the envelope within 1/10 th of the Hill radius.
The black curves are derived with the fiducial opacity, while the green curves are derived with opacity with $a_{max}$=10 cm. We can see that our fiducial luminosity in the simulations ($5.92\times 10^{27}$ ergs/s) is at the higher end of the luminosity curve, which occurs at the early contraction stage or late run-away accretion stage. The reduced luminosity in our simulations ($5.92\times 10^{26}$ ergs/s) is more aligned with the typical envelope luminosity.

The resulting Jupiter envelope evolution is very similar to previous results. The 10 earth mass core can undergo runaway accretion within 10 Myr, and the reduced opacity can speed up the atmosphere accretion. These agreements are not surprising since our 3-D simulations basically confirm the previous 1-D analytical/semi-analytical approach. The improvement we proposed (section \ref{sec:1dana}) does not change the results qualitatively.

\subsection{Chemical Abundances of the Planetary Atmosphere}
\label{sec:chemical}
Although our 3-D simulations produce similar planetary structure and planet evolution as previous 1-D models, the 3-D simulations reveal efficient mass recycling which can change the chemical abundances of the planetary atmosphere.

For a planet with a final atmosphere mass of $m_{atm}$, let's assume that the initially accreted material has a fraction ($f_i$) of this mass in the metal. But the planet is embedded in the disc and the disc material has a fraction ($f_{d}$) of its mass in the metal. After exchanging the mass of $\Delta$m, the metal fraction of the atmosphere becomes
\begin{equation}
    f_{o}=\frac{(m_{atm}-\Delta m)f_i+\Delta m f_d}{m_{atm}}\,.
\end{equation}
If the disc has the same metallicity as the planet ($f_d=f_i$), the atmosphere's chemical abundances won't change. If all the atmosphere is recycled ($\Delta m=m_{atm}$), the atmosphere metallicity will change to the disc's metallicity. Thus, both $\Delta m$ and the difference between $f_i$ and $f_d$ are crucial for the planet's final metallicity. 

From the first-order estimate, $\Delta m$ cannot be higher than the local disc mass. If we assume that the planet can at most accrete the disc material within 1 disc scale height at each side of the planet, this local disc mass is 0.9, 22, 214, 1342 $M_{\oplus}$ if the planet is at 0.1, 1, 5, 20 au based on our adopted disc model in Section \ref{sec:diskstructure}. Thus, there is a larger mass reservoir for recycling at the outer disc.
On the other hand, for a radiative envelope around a 10 $M_{\oplus}$ core, the recycling can penetrate until $\sim$0.1 Hill radius, which corresponds to 5, 51, 253, 1012 $R_{\oplus}$ at these distances. For a planet at 0.1 au, almost all the atmosphere beyond the core radius ($\sim$2 $R_{\oplus}$) can be recycled, while, for a planet that is far away from the star, the recycling process may only exchange a tiny fraction of the atmosphere at its surface. Overall, the planet closer to the star is more affected by disc recycling, although there is a limited amount of disc mass reservoir. Thus, discovered super-Earths or mini-Neptunes within 1 au may have local disc metallicity imprinted no matter where they formed initially in the disc. 

For a planet with an outer convective envelope (e.g. due to the high luminosity from pebble accretion or runaway accretion), the recycling may penetrate to a deeper part of the atmosphere. On the other hand, these high luminosity phases can be short compared with the disc lifetime. 

The other effect, $f_{i}-f_{d}$, is determined by the disc evolution and the planet migration. During planet formation, the chemical abundances of the disc change with time (e.g. \citealt{Li2020} ) so that the initially accreted material may have a different abundance from the later surrounding disc material. At the same time, if the planet migrates in the disc, the planet can exchange with the disc material during its migration. 

It is feasible to incorporate the mass exchange into the core-accretion planet formation model considering disc evolution, planet migration, and atmosphere accretion, and derive the final planet metallicity. But it is beyond the scope of this work, and we will leave it for future studies.

\section{Conclusion}
\label{sec:conclusion}
The traditional model of giant planet formation through core-accretion is derived by the 1-D quasi-static approach. Recently, this 1-D approach has been challenged by 3-D numerical simulations which show complicated flow patterns between the envelope and the disc. Some works suggest that the 
significant recycling between the planetary envelope and the disc can slow or even stall the envelope accretion, potentially explaining the large number of discovered super-Earths and mini-Neptunes. 

To reconcile 1-D isolated envelope models and 3-D disc simulations, we have carried out radiation hydrodynamic simulations for 1-D and 3-D isolated spherical envelopes, and 3-D envelopes embedded in discs. Different from most previous 3-D simulations, we heat the envelopes at specific rates so that the envelopes can achieve steady states, similar to the traditional static models. Furthermore, we have carried out the 3-D isolated sphere simulations to bridge 1-D models and 3-D disc models. When we compare these 3-D isolated sphere simulations with the 1-D isolated sphere simulations, we can understand the role played by convection. When we compare them with the 3-D disc simulations, we can understand the role played by disc recycling. 

We have updated the opacity table for these simulations. Our new table uses the dust opacity derived from protoplanetary disc observations. The molecular and atomic opacities are also updated to cover the $\rho-T$ condition in a forming planet's envelope. Both Rosseland mean and Planck mean opacities for different metalicities have been derived and provided publicly through GitHub.

We test our 1-D radiation simulations against analytical solutions. Insufficient numerical resolution can lead to artificial heating close to the inner boundary where the density gradient is highest. Thus, a resolution study, at least with the 1-D setup, is crucial for the correct envelope simulations.

When the luminosity is capable of driving convection, 3-D simulations are needed since 1-D radiation simulations do not allow the convective motion. With efficient energy transport by convection, the resulting envelope structure follows  $\nabla\sim\nabla_{ad}$, as expected. The convective envelope in the disc has a similar structure as the isolated envelope, except with a slightly higher density due to a lower entropy. 

When we increase the luminosity to make the envelope more convective, the convective motion becomes transonic which reduces the efficiency of convection. The envelope also shows significant asymmetric features since the envelope does not have time to adjust to the transonic convective motion.

When we adjust the parameters (e.g. a lower opacity, a hotter disc, or a lower luminosity) to make the envelopes radiative, we find almost identical structure among 1-D/3D isolated envelopes and 3-D envelopes in discs. We provide a modified 1-D semi-analytical approach which can fully reproduce our 3-D disc simulations for both convective and radiative envelopes. 

Using a passive scalar, we indeed observe significant mass recycling on the orbital timescale. For the radiative envelope, recycling can only penetrate to $\sim$0.1-0.2 Hill radius, while, for the convective envelope, the convective motion can ``dredge up'' the deeper part of the envelope so that the whole convective envelope is recycled efficiently with the disc material. This mass exchange has important implications on the proto-Jupiter's composition. For an example, a migrating proto-Jupiter can quickly (on the orbital timescale) mix the local disc material with its envelope material so that the envelope composition does not reflect where the planet originally formed. 

Although the recycling has large effects on mass exchange, it has limited effects on the envelope's thermal structure, at least for 10 $M_{\oplus}$ planetary cores at 5 au. We estimate various timescales to study the effects of recycling on the envelope structure.
For a convective envelope, the recycling timescale is  similar to the convective timescale. Thus, recycling can quickly exchange the envelope material with the disc material even to the deepest part of the outermost convective envelope. For a radiative envelope, the mass recycling is only efficient beyond 0.1 $r_H$. Our crude estimate suggests that the cooling timescale for the recycled material is shorter/comparable to the recycling timescale, which may explain the limited effects of recycling on the thermal structure of the envelope. 

With the updated opacity table, equation of states, and 1-D models, we calculate Jupiter's atmosphere accretion with a 10 $M_{\oplus}$ core, and confirm that it can undergo runaway accretion within the disc's lifetime using our fiducial opacity. With a lower opacity in the envelope (opacity from 10 cm grains), the time to runaway accretion can be shorter than a Myr. 

 {
Finally, we discuss how the recycling process can affect the chemical abundances of the planet atmosphere. Both the amount of the recycled mass and the difference between the envelope metallicity and the disc metallicity are key for determining the final planet's metallicity. For a planet close to the star, the recycling process can efficiently exchange the planet's gaseous atmosphere with the disc, although there is a limited amount of disc mass reservoir at the inner disc for the exchange. Overall, the discovered super-Earths or mini-Neptunes within 1 au may have local disc metallicity imprinted no matter where these planets formed initially in the disc. }

\section*{Acknowledgments}
The authors thank the referee for a very helpful report, especially regarding the implications on chemical abundances of the planetary atmosphere. 
This research was supported by NASA TCAN award 80NSSC19K0639. All  simulations are carried out using computer supported by the Texas Advanced Computing Center (TACC) 
at The University of Texas at Austin through XSEDE grant TG-AST130002 and  from the NASA High-End Computing (HEC) program through the NASA Advanced Supercomputing (NAS) Division at Ames Research Center.  Z. Z. acknowledges support from the National Science Foundation under CAREER Grant Number AST-1753168. 
The Center for Computational Astrophysics at the Flatiron Institute is
supported by the Simons Foundation.  ANY acknowledges support from NASA by grant NNX17AK59G

\section*{DATA AVAILABILITY}
The data underlying this article are available in the article and in its online supplementary material.

\bibliographystyle{mnras}
\input{msMNRAS3.bbl}

\appendix
\onecolumn
\section{Source Terms under the Rotating Frame}
\label{sec:source}
Since we solve the fluid equations in the rotating frame which rotates around the central star at an angular frequency of $\Omega_0$,
we need to add the Coriolis force ($-2\bm{\Omega_0}\times\bm{V}$) and the centrifugal force ($-\bm{\Omega_0}\times(\bm{\Omega_0}\times\bm{r})$) as external source terms.
We want to add these two terms in a way to: 1) cancel out the central star's gravitational force and the geometrical source terms as much as possible, and 2) conserve  angular momentum in
the inertial frame. Since the gravitational force from the central star is the largest force in the system, any numerical imbalance between the gravitational force term and the other terms can lead to 
strong perturbations to a steady fluid. Besides the radial force balance, conserving the angular momentum is crucial for accurately simulating a disc in a rotating frame \citep{Kley1998}.

Most of the derivation below is applicable to different hydrodynamical codes, although some are specifically for the \texttt{Athena++} code which we will state explicitly. 
For the radial momentum equation, the radial Coriolis and centrifugal forces are
$F_r=2\Omega_0 v_{\phi} {\rm sin}\theta+\Omega_0^2 r {\rm sin}^2 \theta$. For  a Keplerian flow, this force needs to balance the gravitational force and the geometric source term $v_{\phi}^2/r$.
In \texttt{Athena++}, the gravitational force of a point mass is implemented as
\begin{equation}
F_{*}=src1i_i\frac{GM}{r_i}\,,
\end{equation}
where 
\begin{equation}
src1i_i=\frac{1/2\left(R_{i+1/2}^2-R_{i-1/2}^2\right)}{1/3\left(R_{i+1/2}^3-R_{i-1/2}^3\right)}\,.
\end{equation}
Thus, we implement the radial source terms from the Coriolis and centrifugal forces using the same quantities
\begin{equation}
\rho F_r=\Omega_0 r_i {\rm sin}\theta_j\left(F_{\phi,k+1}+F_{\phi,k}\right)\times src1i_i+\rho\Omega_0^2r_i^2{\rm sin}^2\theta_{j}\times src1i_i\,.\label{eq:Fr}
\end{equation}
where $F_{\phi,k}$ is the Riemann flux at the constant $\phi_k$ interface for the density continuity equation.  The first term
on the right side is the Coriolis force, while the second term is the centrifugal force.

Similarly, for the $\theta$ momentum equation, the Coriolis and centrifugal forces are
$F_\theta=2\Omega_0 v_{\phi} {\rm cos}\theta+\Omega_0^2 r {\rm sin}\theta{\rm cos}\theta$. This force needs to balance the geometric source term (cot$\theta$ $v_{\phi}^2$/r) which uses the factor of
$src1i_i\times src1j_j$ in \texttt{Athena++} where 
\begin{equation}
src1j_j=\frac{{\rm sin}\theta_{j+1}-{\rm sin}\theta_{j}}{{\rm cos}\theta_{j+1}-{\rm cos}\theta_{j}}\,.
\end{equation}
Thus, we implement the $\theta$ source terms using
\begin{equation}
\rho F_{\theta}=\Omega_0 r_i {\rm sin}\theta_j\left(F_{\phi,k+1}+F_{\phi,k}\right)\times src1i_i\times src1j_j+\rho\Omega_0^2r_i^2{\rm sin}^2\theta_{j}\times src1i_i\times src1j_j\,.\label{eq:Ftheta}
\end{equation}
In both Equations \ref{eq:Fr} and \ref{eq:Ftheta}, we use the Riemann flux to represent $\rho v_{\phi}$ to slightly improve the accuracy. 

For the $\phi$ momentum equation, we can absorb both the Coriolis and centrifugal terms into the divergence operator so that it is written as a conservative form \citep{Kley1998},
\begin{eqnarray} 
    \frac{\partial \left(\rho \left(v_{\phi}+\Omega_0 r\sin\theta\right)r\sin\theta\right)}{\partial t} + \nabla\cdot \left(r\sin\theta \rho \bold{u}\left(v_{\phi}+\Omega_0 r\sin\theta\right)\right)=0 \,.\label{eq:azimuthalmomentumco2}
\end{eqnarray}
If we solve Equation \ref{eq:azimuthalmomentumco2} directly using the finite volume method, we can conserve the total angular momentum ($\rho(v_\phi+\Omega_0 r_c {\rm sin}\theta_c)  r_c {\rm sin}\theta_c$) 
to the machine precision, where $r_c$ and $\theta_c$ represent the distance and angle of each cell center. Unfortunately, solving Equation \ref{eq:azimuthalmomentumco2} directly can significantly change the structure
of the code. We would still like the code to solve the same set of equations no matter what coordinate systems we are using. Thus, we expand Equation  \ref{eq:azimuthalmomentumco2} into the discretized form and separate the terms which are associated with the Coriolis and centrifugal forces. The time derivative of the density is replaced by the discretized continuity equation. In this way,
adding the derived source term is equivalent to solving Equation \ref{eq:azimuthalmomentumco2} directly while keeping the structure of the code intact.  The derived source term is
\begin{eqnarray}
\rho F_{\phi}&=&-\frac{\Omega_0 {\rm sin}\theta_c\left({\rm cos}\theta_j-{\rm cos}\theta_{j+1}\right)\Delta \phi }{r_c \Delta V}\left(r_{i+1}^4 F_{r,i+1}-r_c^2 r_{i+1}^2 F_{r,i+1}-r_{i}^4 F_{r,i}+r_c^2 r_{i}^2 F_{r,i}\right)\nonumber\\
&&-\frac{\Omega_0 r_c \left(r_{i+1}^2-r_i^2\right)\Delta\phi}{2{\rm sin}\theta_c \Delta V}\left({\rm sin}^3\theta_{j+1}F_{\theta,j+1}-{\rm sin}^3\theta_{j}F_{\theta,j}-{\rm sin}^2\theta_c{\rm sin}\theta_{j+1}F_{\theta,j+1}+{\rm sin}^2\theta_c{\rm sin}\theta_{j}F_{\theta,j}\right)\,,\label{eq:Fphi}
\end{eqnarray}
where $\Delta V$=$1/3(r_{i+1}^3-r_{i}^3)({\rm cos}\theta_j-{\rm cos}\theta_{j+1})\Delta\phi$.
We chose $r_c=(r_{i+1}+r_{i})/2$ and sin$\theta_c=({\rm sin}\theta_{j+1}+{\rm sin}\theta_{j})/2$, so that Equation \ref{eq:Fphi} becomes
\begin{eqnarray}
\rho F_{\phi}&=&-\frac{3\Omega_0\left( {\rm sin}\theta_{j+1}+ {\rm sin}\theta_j\right)\left(r_{i+1}^2 F_{r,i+1}\left(3r_{i+1}+r_i\right)+r_i^2F_{r,i}\left(3r_i+r_{i+1}\right)\right)}{4\left(r_{i+1}+r_i\right)\left(r_{i+1}^2+r_{i+1}r_i+r_i^2\right)}\nonumber\\
&-&\frac{3\Omega_0 \left(r_{i+1}+r_{i}\right)^2\left( {\rm sin}\theta_{j+1}- {\rm sin}\theta_j\right)\left(\left(3{\rm sin}\theta_{j+1}+{\rm sin}\theta_j\right){\rm sin}\theta_{j+1}F_{\theta,j+1}+\left(3{\rm sin}\theta_j+{\rm sin}\theta_{j+1}\right){\rm sin}\theta_j F_{\theta,j}\right)}{8\left( {\rm sin}\theta_{j+1}+ {\rm sin}\theta_j\right)\left( {\rm cos}\theta_{j}- {\rm cos}\theta_{j+1}\right)\left(r_{i+1}^2+r_{i+1}r_i+r_i^2\right)}\,.
\end{eqnarray}
Since the Coriolis force does not do any work, we only add the work done by the centrifugal force in the energy equation. Again, we use the density flux (F) instead of $\rho v$ to calculate the work done by the centrifugal force.

\section{Constant Radiation Flux Boundary Condition }
\label{sec:constflux}
Our 1-D test problem requires us to have a constant radiation flux coming out of the inner radial boundary. 
Since we solve the specific intensity for the radiative transfer, we need to provide the specific intensity in all directions within the ghost zones. 
To maintain a constant radiation flux at the inner boundary, the second momentum equation of the radiative transfer equation
\begin{equation}
\nabla\cdot{\bm P_{r,c}}=-\sigma_{T,c}{\bm F_{r,c}}\,,
\end{equation} 
where ${\bm P_{r,c}}$ and ${\bm F_{r,c}}$ are the radiation pressure tensor and the radiation flux in the code unit, and
$\sigma_{T,c}$ is the total opacity in the code unit, indicates that
\begin{equation}
E_{r,c}(r_g)=T_{in}^4+3\sigma_{T,c}\left(r_{in}-r_g\right)F_{r,c}\,,\label{eq:Erc}
\end{equation}
where $r_g$ and $r_{in}$ are the radial position of each ghost zone and the first active zone, and we have assumed the Eddington approximation
$P_{r,c}=E_{r,c}/3$ and $E_{r,c}=T_c^4$ for the LTE condition. 

Using two stream approximation, we assume that all rays pointing to the negative $r$ direction have the same intensity $I_-$, while all rays pointing
outwards have the intensity $I_+$. Then, we have
\begin{eqnarray}
E_{r,c}&=&I_-\times \sum_- w_-+I_+\times\sum_+ w_+\nonumber\\
F_{r.c}&=&I_-\times \sum_- w_-\mu_{r-}+I_+\times\sum_+ w_+\mu_{r+}\,.
\end{eqnarray}
where $w$ and $\mu_r$ are the weights and r-direction cosines as defined in \cite{Davis2012}, and $\sum_-$ or $\sum_+$ is the summation of all rays in negative or positive $r$ directions. The sign of - and + represent rays pointing to the negative or positive $r$ direction. 
If we define $\sum_- w_-$ as a$_-$, $\sum_+ w_+$ as a$_+$, $\sum_- w_-\mu_{r-}$ as b$_-$, and $\sum_+ w_+\mu_{r+}$ as b$_+$, we can solve for $I_-$ and $I_+$ as
\begin{eqnarray}
I_-&=&\frac{E_{r,c}/a_+-F_{r,c}/b_+}{a_-/a_+-b_-/b_+}\nonumber\\
I_+&=&\frac{E_{r,c}/a_--F_{r,c}/b_-}{a_+/a_--b_+/b_-}\,.
\end{eqnarray}
Using $E_{r,c}$ calculated in Equation \ref{eq:Erc}, we assign $I_-$ and $I_+$ for rays propagating in the negative and positive $r$ directions for every cell in the ghost zones.

\bsp
\label{lastpage}
\end{document}